\documentclass[twocolumn]{aastex63}
\usepackage{graphicx,natbib,color,bm,url,times}
\usepackage{amsmath}
\graphicspath{{./fig/}{./png/}}
\newcommand{\EQ}{\begin{equation}}
\newcommand{\EN}{\end{equation}}
\newcommand{\EQL}{\begin{align}}
\newcommand{\ENL}{\end{align}}
\newcommand{\EQA}{\begin{eqnarray}}
\newcommand{\ENA}{\end{eqnarray}}
\newcommand{\eq}[1]{Eq.~(\ref{#1})}
\newcommand{\EEq}[1]{Equation~(\ref{#1})}
\newcommand{\eqs}[2]{Eqs.~(\ref{#1}) and~(\ref{#2})}
\newcommand{\eqss}[2]{Eqs.~(\ref{#1})--(\ref{#2})}

\newcommand{\App}[1]{Appendix~\ref{#1}}

\newcommand{\Sec}[1]{Sect.~\ref{#1}}

\newcommand{\Fig}[1]{Figure~\ref{#1}}

\newcommand{\bra}[1]{\langle #1\rangle}

{}
{}
\newcommand{\meanFFFF}{\overline{\mbox{\boldmath ${\cal F}$}}{}}{}
\newcommand{\meanemf}{\overline{\cal E} {}}

{}
{}
\newcommand{\meanEMF}{\overline{\mbox{\boldmath ${\cal E}$}}{}}{}
{}
{}
{}
{}
{}
{}
{}
{}
{}
{}
{}
{}
{}
{}
\newcommand{\meanUU}{\overline{\bm{U}}}

{}

\newcommand{\meanA}{\overline{A}}
\newcommand{\meanB}{\overline{B}}

\newcommand{\meanG}{\overline{G}}

\newcommand{\meanJ}{\overline{J}}

\newcommand{\meanFFF}{\overline{\cal F}}

{}

{}
{}
{}
\newcommand{\calA}{{\cal A}}{}

\newcommand{\nnn}{\hat{\bm{n}}}

\newcommand{\xxx}{\hat{\mbox{\boldmath $x$}} {}}
\newcommand{\yyy}{\hat{\mbox{\boldmath $y$}} {}}

\newcommand{\meanAA}{{\overline{\bm{A}}}}
\newcommand{\meanBB}{{\overline{\bm{B}}}}
\newcommand{\meanJJ}{{\overline{\bm{J}}}}

\newcommand{\aaaa}{\bm{a}}
\newcommand{\jj}{\bm{j}}

\newcommand{\bb}{\bm{b}}

\newcommand{\bzz}{b^{(00)}}

\newcommand{\BB}{\bm{B}}

\newcommand{\JJ}{\bm{J}}

\newcommand{\AAA}{\bm{A}}

\newcommand{\UU}{\bm{U}}
\newcommand{\UUS}{\bm{U}^{(S)}}
\newcommand{\FF}{\bm{F}}
\newcommand{\FFK}{\bm{F}_{\rm K}}
\newcommand{\FFM}{\bm{F}_{\rm M}}
\newcommand{\ffK}{\bm{f}_{\rm K}}
\newcommand{\ffM}{\bm{f}_{\rm M}}

\newcommand{\uu}{\bm{u}}

\newcommand{\nab}{{\bm{\nabla}}}

\newcommand{\OO}{\bm{\Omega}}

\newcommand{\SSSS}{\mbox{\boldmath ${\sf S}$} {}}

\newcommand{\ii}{{\rm i}}

\newcommand{\DDD}{{\cal D} {}}

\newcommand{\const}{{\rm const}  {}}

\def\degr{\hbox{$^\circ$}}

\def\H{\mbox{\rm H}}

\def\Ma{\mbox{\rm Ma}}

\def\ShK{\mbox{\rm Sh}_{\rm K}}

\def\St{\mbox{\rm St}}

\def\Pm{\mbox{\rm Pr}_{\rm M}}
\def\Rm{\mbox{\rm Re}_{\rm M}}
\def\Rey{\mbox{\rm Re}}

\def\Lu{\mbox{\rm Lu}}

\def\cs{c_{\rm s}}

\def\kf{k_{\rm f}}

\def\Brms{B_{\rm rms}}

\def\urms{u_{\rm rms}}

\def\etat{\eta_{\rm t}}

\def\etaT{\eta_{\rm T}}

\newcommand{\W}{\,{\rm W}}
\newcommand{\V}{\,{\rm V}}

\newcommand{\T}{\,{\rm T}}
\newcommand{\G}{\,{\rm G}}

\newcommand{\K}{\,{\rm K}}
\newcommand{\g}{\,{\rm g}}
\newcommand{\s}{\,{\rm s}}

\newcommand{\m}{\,{\rm m}}

\newcommand{\J}{\,{\rm J}}

\newcommand{\A}{\,{\rm A}}

\def\apjs{ApJS}

\def\QKTFM{{\sc qktfm}}

\begin{document}

\correspondingauthor{Maarit J. K\"apyl\"a}
\email{maarit.kapyla@aalto.fi}
\author[0000-0002-9614-2200]{Maarit J. K\"apyl\"a}
\affiliation{Department of Computer Science, 
	      Aalto University, P.O. Box 15400, FI-00076 Aalto, Finland}
\affiliation{Max Planck Institute for Solar System Research,
              Justus-von-Liebig-Weg 3, D-37077 G\"ottingen, Germany}	      
\affiliation{Nordita, KTH Royal Institute of Technology and Stockholm University, 
              Roslagstullsbacken 23, SE-10691 Stockholm, Sweden}

\author{Javier \'Alvarez Vizoso}
\affiliation{Max Planck Institute for Solar System Research,
              Justus-von-Liebig-Weg 3, D-37077 G\"ottingen, Germany}	
              
\author{Matthias Rheinhardt}
\affiliation{Department of Computer Science, 
	      Aalto University, PO Box 15400, FI-00076 Aalto, Finland}

\author[0000-0002-7304-021X]{Axel Brandenburg}
\affiliation{Nordita, KTH Royal Institute of Technology and Stockholm University, 
              Roslagstullsbacken 23, SE-10691 Stockholm, Sweden}
\affiliation{Department of Astronomy, AlbaNova University Center,
              Stockholm University, SE-10691 Stockholm, Sweden}
\affiliation{McWilliams Center for Cosmology \& Department of Physics, Carnegie Mellon University, Pittsburgh, PA 15213, USA}

\author[0000-0001-6097-688X]{Nishant K. Singh}
\affiliation{Inter-University Centre for Astronomy and Astrophysics,
 Post Bag 4, Ganeshkhind, Pune 411 007, India}
\affiliation{Max Planck Institute for Solar System Research,
              Justus-von-Liebig-Weg 3, D-37077 G\"ottingen, Germany}	      

\title{On the existence of shear-current effects in magnetized burgulence}
\shorttitle{Shear-current effect in magnetized burgulence}
\shortauthors{K\"apyl\"a et al.}

\date{\!$ \, $Revision: 1.460 $ $\!}

\begin{abstract}
The possibility of explaining shear flow dynamos by a magnetic
shear-current (MSC) effect is examined via numerical simulations.
Our primary diagnostics is the determination of the turbulent magnetic
diffusivity tensor $\boldsymbol{\eta}$.
In our setup, a negative sign of its component $\eta_{yx}$
is necessary for coherent dynamo action by the SC effect.
To be able to measure turbulent transport coefficients from systems with 
magnetic background turbulence, we present an extension of the 
test-field method (TFM) applicable to our setup where the pressure
gradient is dropped from the momentum equation: the nonlinear TFM (NLTFM). 
Our momentum equation is related to Burgers' equation and the
resulting flows are referred to as magnetized burgulence.
We use both stochastic
kinetic and magnetic forcings to mimic cases without and with 
simultaneous small-scale dynamo action.
When we force only kinetically, negative $\eta_{yx}$ are obtained with 
exponential growth in both the radial and azimuthal 
mean magnetic field components.
Using 
magneto-kinetic
forcing,  
the field growth is no longer exponential, while NLTFM yields
positive $\eta_{yx}$. 
 By employing an alternative forcing from which
wavevectors 
whose components correspond to the largest scales
are removed,
the exponential growth is
recovered, but the NLTFM results do not change significantly.
Analyzing the dynamo excitation conditions
for the coherent SC and incoherent $\alpha$ and SC effects
shows that the incoherent effects are the main drivers of the dynamo
in the majority of cases. 
We find no evidence for MSC-effect-driven dynamos in our simulations.
\end{abstract}

\section{Introduction}
In recent years, 
there has been a lot of interest in 
the possibility of large-scale
dynamo (LSD) action through the shear-current effect 
 \citep{RK03,RK04}
in flows where more conventional dynamo effects, such as the $\alpha$ effect
arising through stratification and rotation,
cannot operate.
In turbulence lacking helicity
due to, say,  the absence of rotation
or stratification in density or turbulence intensity, 
the $\alpha$ tensor 
vanishes.
The turbulent magnetic diffusivity tensor
$\boldsymbol{\eta}$,
however, is always
found
to have finite and positive diagonal components.
Its off-diagonal components are in general also finite
if there is rotation or shear.
Rotation alone gives rise to the
$\OO\times\JJ$ or R\"adler   
effect \citep{Rae69,Rae69b}
and shear alone to
the shear-current (SC) effect.
For a suitable sign of the relevant
off-diagonal component of  
$\boldsymbol{\eta}$, the latter 
effect can
lead to dynamo action even
without rotation,
but the former would not do so
without shear.
Both the R\"adler and SC effects have been discussed as additional or
even major dynamo effects in stars \citep{PS09}, accretion
disks \citep{LO08,Bla10}, and galactic magnetism \citep{CS18}.

Astrophysical flows are also subject to vigorous small-scale
dynamo (SSD) action, which should occur in any flow where 
the magnetic Reynolds and Prandtl numbers are large enough. 
The SSD produces strong, fluctuating magnetic fields
at scales smaller than the forcing scale of the turbulence,
on time scales short in comparison to the LSD instability
\citep[see, e.g.,][]{BSK12}.
Usually, the SSD is thought to be detrimental to 
$\alpha$-effect-driven dynamos,
where dynamo action can be strongly suppressed in regimes with high 
magnetic
Reynolds number
\citep[e.g.,][]{CV91,VC92},
unless the system can get a rid of small-scale magnetic helicity
by interacting with its surroundings through helicity fluxes
\citep[e.g.,][]{BF01,Axel01,BS05}.
In the absence of 
magnetic background turbulence 
it has not yet been possible to verify the
existence of a dynamo driven by the 
SC effect 
\citep[][]{BRRK08,You08,SJ15}.
Failure to understand the origin of large-scale magnetic fields in
these numerical works in terms of the SC effect, together with the findings
of significant $\alpha$ fluctuations in \cite{BRRK08}, provided
enough motivation to explore the possibility of 
LSD action driven solely by a
fluctuating $\alpha$ in shearing
systems. Such an incoherent $\alpha$-shear dynamo was studied analytically
in a number of previous works, suggesting a possibility of generation
of large-scale magnetic fields due to purely temporal fluctuations
in $\alpha$ in the presence of shear \citep{HMS11,MB12,SS14}.

It has been claimed, however, that
in the presence of forcing in the induction equation, mimicking 
magnetic background turbulence provided, e.g., by the SSD, a
thus {\it magnetically} driven SC dynamo effect
 exists \citep[][]{SB15a,SB15b,SB16}.
In their analytic study, under the second-order correlation approximation (SOCA),
\cite{SB15a} argued for a significant magnetic contribution of the type
leading to coherent dynamo action in systems with
both
shear and rotation
with typical $q=-S/\Omega$ values for galactic or accretion
disks, while in the regime of shear dominating over rotation, relevant
to our current study, such contribution was found to be weak.
In \cite{SB16}, it was furthermore argued that 
the magnetic shear-current (MSC) effect
``arises exclusively from the pressure response of the velocity fluctuations."
As we demonstrate in Appendix~\ref{analyticetayx}, based on an
analytical calculation, a magnetic
contribution to $\eta_{yx}$ exists even when the pressure term
is dropped, but 
then it likely has
a sign that is unfavorable for dynamo action. However, since 
these analytic results suffer from many simplifications, 
they cannot provide a conclusive picture. Hence,
it is important to study this issue numerically, which is one of our
aims in this 
work.

In their numerical studies, \cite{SB15b,SB16}
reported the generation of a large-scale magnetic field,
usually on the scale of the computational domain, 
with magnetic forcing,
while in the case of kinetic forcing only, the generated
patterns were reported to be temporally more erratic and spatially less coherent.
For a flow in the
$y$ direction, sheared in $x$,
an attempt was made to measure the turbulent transport coefficients using the
second-order cumulant expansion 
method of \cite{Marston08}, and the results
indicated negative $\eta_{yx}$ and $\eta_{xx}$ in the presence of magnetic forcing
\citep{SB15b}.
Incidentally, if confirmed in this case, a negative $\eta_{xx}$ 
could
also imply dynamo action \citep{Lan+99,DBM13}.
At that time, however, a suitable test-field method (TFM), 
providing another measurement tool for
the turbulent transport coefficients, was not yet available. 

Here, we present first steps toward
such a toolbox, 
extending the method developed by 
\citet[][hereafter RB10]{RB10}
to include the self-advection term
and rotation, albeit still limited to simplified MHD 
(SMHD)
equations, with the pressure gradient
term being dropped. Although this method does not yet provide a completely suitable tool
for the systems studied by \cite{SB15b,SB16}, it does provide a working 
solution for simplified shear dynamos 
with magnetic forcing,
mimicking SSD, and can be envisioned to enable
important scientific insights. 
In this paper, we present the method,
referred to as the
``nonlinear test-field method" (NLTFM),
and tests against previously studied cases, 
along with other validation results. 
As our major topic, we
analyze runs with SMHD
equations that exhibit dynamo action
in the same parameter regime as previously claimed to host 
MSC-effect dynamos.

\section{Model and Methods}

We perform local Cartesian box simulations with shearing-periodic boundary
conditions to implement large-scale shear as a linear background flow imposed on the 
system. 
The shear occurs in the $x$ direction, which 
could represent, e.g., the direction 
from the rotational center of a cosmic body. Here, $y$ is the streamwise,
or azimuthal, direction, and $z$ points into the vertical direction.
The magnitude of the shearing motion is described by the input parameter $S$ such that the imposed linear shear flow is
${\bm U}^{\rm S} =  S x \bm{\hat{y}}.$
The rotation of the domain, ${\bm \Omega}=(0,0,\Omega)$, is described by 
the input parameter $\Omega$, the magnitude of the angular velocity.
In the
simulations reported in this paper, however, rotation is 
neglected, as here we concentrate on studying 
the possibility of the SC effect alone.
We will, however, retain rotation in 
the model equations
for completeness.
Our boxes have edge lengths 
$L_x=L_y$, and $L_z$ with aspect ratio 
$\calA=L_z/L_x$
chosen equal to one in many
cases, but we consider also vertically elongated boxes 
with  $\calA=4,8,16$.
All calculations were carried out with the {\sc Pencil Code}\footnote{\url{http://github.com/pencil-code}}, 
see e.g. \cite{PC2020}.

\subsection{Simplified MHD}

As stated in the introduction, the equations of 
SMHD
as defined here are similar to
those of MHD,
but lack the
pressure gradient.
Correspondingly, the density $\rho$ is held constant.
We solve the equations for the magnetic vector potential $\AAA$
and the velocity $\UU$,
\EQA
\DDD^A \AAA&=&\UU\times\BB+\FFM  +\eta\nab^2\AAA,\label{dAA} \\
\DDD^U\UU &=&- \UU\cdot\nab \UU + \JJ\times\BB/\rho+\FFK  \label{dUU} \nonumber \\
                       &&+\nu(\nab^2\UU + \nab \nab\cdot\UU/3)
\ENA
with the linear expressions
\EQA
\DDD^{A} \AAA&=&\DDD\AAA+S A_y \xxx, \\
\DDD^U \UU&=& (\DDD+2\OO\times)\UU +S U_x \yyy, \\
\cal D &=& \partial_t + S x \partial_y.
\ENA
$\BB=\nab\times\AAA$ is the magnetic field, $\JJ=\nab\times\BB$
is the current density in units where the vacuum permeability is unity,
$\FFK$ and $\FFM$ are kinetic and magnetic forcing functions, respectively,
$\eta$ is the (molecular) magnetic diffusivity, and $\nu$ is the kinematic viscosity,
both assumed constant.
\EEq{dUU} can be considered a three-dimensional 
generalization of Burgers' equation,
which is why one refers to its turbulent solutions as ``burgulence";
see the review by \cite{FB00} on such flows.

The main advantage of using SMHD
is to
avoid the necessity of dealing with density fluctuations  and corresponding effects in the mean quantities.
However,
as self-advection $\UU\cdot\nab \UU$ is no longer discarded, we are here more general than RB10,
the models of which suffered, in physical
terms,
from the implied assumption of slow fluid motions,
that is, small Strouhal numbers ($\St\ll1$) or 
Reynolds numbers ($\Rey\ll1$).
Completely neglecting
the self-advection term is inadequate in the present context given that shear plays its
essential role just via this term. So merely the terms arising from an additional mean flow and
from the fluctuating velocity alone could be dropped.
Neglecting the latter, 
however, would be equivalent to restricting the
method to SOCA with respect
to the self-advection term, which is not desirable.

\subsection{Full MHD}

The full MHD (FMHD)
system of equations,
here with an isothermal equation
of state, is more complex because
of the occurrence of the pressure gradient, as a result of
which we need
an additional
evolution equation for the density.
Also the viscous force is more complex, hence
\EQA
\DDD^{A}\AAA&=&\UU\times\BB+\FFM+\eta\nab^2\AAA,\nonumber\\   
\rho(\DDD^{U} +\UU\cdot\nab)\UU+\nab p&=&\JJ\times\BB+\rho \FFK+\nab\cdot(2\nu\rho\SSSS),\nonumber\\
(\DDD+\UU\cdot\nab)\ln\rho&=&-\nab\cdot\UU\label{dlr2}.
\ENA
Here, ${\sf S}_{ij}=(U_{i,j}+U_{j,i})/2-\delta_{ij}\nab\cdot\UU/3$   
are the
components of the rate-of-strain tensor $\SSSS$, where commas denote
partial differentiation, 
and $p$ is the pressure related to the density
via $p=\cs^2\rho$, with $\cs=\const$ being the isothermal sound speed.

\subsection{Test-field methods}

Throughout, we define mean quantities by horizontal averaging, i.e., averaging over $x$ and $y$, denoted by an overbar. 
So 
the means
depend on $z$ and $t$ only.
Fluctuations are denoted by lowercase symbols or a prime, e.g.,
$\aaaa=\AAA-\meanAA$, $\uu=\UU-\meanUU$,  $(\uu\times\bb)'=\uu\times\bb-\overline{\uu\times\bb}$,
and $\bm{f}_{\rm K,M}=\bm{F}_{\rm K,M}-\overline{\bm F}_{\rm K,M}$.
The horizontal average is
normally taken to 
obey the Reynolds rules.
In situations with linear overall shear though,
the complication arises that
$\UU^S \ne \overline{\UU^S}$ (when $\UU^S$ is defined to be $\propto x$, the mean even vanishes),
being hence not a pure mean,
while ${\partial_i U^S_j}$ is spatially constant, hence a pure mean.
So the Reynolds rule ``averaging commutes with differentiating" is violated.
However, $(\UU^S\cdot\nab G)'=\UU^S \cdot\nab g$
for an arbitrary quantity $G=\meanG+g$. This is a consequence of
$\UU^S \cdot\nab \meanG = 0$ and $\overline{\UU^S \cdot\nab g} = \int\! \int Sx \,\partial_y g\, dx dy = \int Sx \left(\int \partial_y g \,dy \right) dx =0$, the latter 
because of periodicity in $y$.
Thus, $\UU^S$ can effectively be treated as a mean flow.

The evolution equations for the fluctuations of the
magnetic vector potential, 
$\aaaa$, and the velocity, $\uu$,
follow from \eqs{dAA}{dUU} as
\EQA
\DDD^{A}\aaaa&=&\uu\times\meanBB+\meanUU\times\bb+(\uu\times\bb)'+\ffM+\eta\nab^2\aaaa,\label{daT}\\
\DDD^{U}\uu  &=&\left(\,\meanJJ\times\bb+\jj\times\meanBB+(\jj\times\bb)' \right)/\rho \nonumber\\ && -(\uu\cdot\nab \uu)' 
 - \meanUU\cdot\nab\uu -  \uu\cdot\nab\meanUU \label{duT} \\
 &&+\ffK + \nu\!\left(\nab^2\uu +\nab \nab\cdot\uu/3\right)\nonumber.
\ENA

\subsubsection{Nonlinear TFM}

In the quasi-kinematic test-field method
(\QKTFM) (see Sect.~\ref{sec:qktfm}), 
the mean electromotive force $\meanEMF = \overline{\bm{u} \times \bm{b}}$
is
a functional of only $\uu$, $\meanUU$, and $\meanBB$ (linear in $\meanBB$).  
However,
in the more general case with a magnetic background turbulence,
this is no longer so.
To deal with this difficulty, RB10
added the  evolution equations for the background turbulence
($\uu_0$,$\bb_0$)
which are similar to \eqs{daT}{duT},
but for zero mean field, to the equations of the TFM.
In general, $\meanEMF=\overline{\uu\times\bb}$ can be split  into a contribution
$\meanEMF_0=\overline{\uu_0\times\bb_0}$ that is independent of the mean field
and
\EQ
\meanEMF_{\bm{\bar{B}}}=\overline{\uu_0\times\bb_{\bm{\bar{B}}}}
+\overline{\uu_{\bm{\bar{B}}}\times\bb_0}
+\overline{\uu_{\bm{\bar{B}}}\times\bb_{\bm{\bar{B}}}},  \label{emfB}
\EN
where $\uu_{\bm{\bar{B}}}$ and $\bb_{\bm{\bar{B}}}$ denote the solutions of \eqs{daT}{duT}
without
forcing (called ``test problems")
which are supposed to vanish for vanishing $\meanBB$.
Thus,
$\uu=\uu_0+\uu_{\bm{\bar{B}}}$,
$\bb=\bb_0+\bb_{\bm{\bar{B}}}$.  $\meanEMF_{\bm{\bar{B}}}$ can  be written in
two equivalent ways as
\EQ
\meanEMF_{\bm{\bar{B}}}=\overline{\uu\times\bb_{\bm{\bar{B}}}}
+\overline{\uu_{\bm{\bar{B}}}\times\bb_0}
=\overline{\uu_0\times\bb_{\bm{\bar{B}}}}
+\overline{\uu_{\bm{\bar{B}}}\times\bb}.
\label{meanEMFBbar}
\EN
Both become linear in quantities with subscript $\meanBB$
when $\bb$ and $\uu$ are identified with the fluctuating fields in the 
``main run",  which is the system (\ref{dAA})--(\ref{dUU}) solved simultaneously
with the 
test 
problems.
In this way, we have recovered the mentioned linearity property of $\meanEMF[\meanBB]$ of the {\QKTFM}.
Likewise, one writes that part of the mean ponderomotive force 
$\meanFFFF_{\bm{\bar{B}}}$,
which results from the Lorentz force as
\EQ
\overline{\jj\times\bb_{\bm{\bar{B}}}}
+\overline{\jj_{\bm{\bar{B}}}\times\bb_0}
\quad\text{or}\quad
\overline{\jj_0\times\bb_{\bm{\bar{B}}}}
+\overline{\jj_{\bm{\bar{B}}}\times\bb}
\label{meanFFFBbar1}
\EN
and that resulting from self-advection as
\EQ
-\overline{\uu\cdot\nab\uu_{\bm{\bar{B}}}}
-\overline{\uu_{\bm{\bar{B}}}\cdot\nab\uu_0}
\quad\text{or}\quad
-\overline{\uu_0\cdot\nab\uu_{\bm{\bar{B}}}}
-\overline{\uu_{\bm{\bar{B}}}\cdot\nab\uu}\,;
\label{meanFFFBbar2}
\EN
see \ Eqs.~(29) and (30) of RB10.
Corresponding expressions can be established  for the fluctuating parts of the bilinear terms,
$(\uu\times\bb)'$, $(\jj\times\bb)'$, and $(\uu\cdot\nab \uu)'$,
occurring in Eqs.~\eqref{daT} and \eqref{duT}.
We recall that these different formulations 
result in different stability properties of the test problems; see also the test results
presented in Appendix~\ref{sect:variants}.
Here, we chose to use
in \eqss{meanEMFBbar}{meanFFFBbar2},
and in the aforementioned expressions for the fluctuating parts of the bilinear terms,
the respective first version, that is,
$\meanEMF_{\bm{\bar{B}}}=\overline{\uu\times\bb_{\bm{\bar{B}}}}
+\overline{\uu_{\bm{\bar{B}}}\times\bb_0}$,
$(\uu\times\bb)' = (\uu\times\bb_{\bm{\bar{B}}})' + (\uu_{\bm{\bar{B}}}\times\bb_0)'$ etc.
This choice forms
what is called the {\sf ju} method; see Table~1 of
RB10.{\footnote{The methods are named after the
fluctuating fields, which are taken over from
the main run; thus the four possible combinations of the expressions in
\eq{meanEMFBbar} and \eq{meanFFFBbar1} yield {\sf ju}, {\sf jb}, {\sf bb}, and {\sf ub}.
Including \eq{meanFFFBbar2} would produce more combinations with three-letter names like {\sf juu} etc.
}

The given alternative formulations 
become equivalent when the mean quantities, possibly evolving  in the main run,
are too weak to have a marked influence on the fluctuating fields.
Then, $\uu \rightarrow \uu_0$ and $\bb \rightarrow \bb_0$,
which defines the kinematic limit.
Employing this
means dropping  terms like $\overline{\uu_{\bm{\bar{B}}}\times\bb_{\bm{\bar{B}}}}$ in mean EMF
and mean force, which is the correct way to obtain the latter
as quantities of first order in $\meanBB$.
Then all possible versions of the NLTFM (which actually ceases to be nonlinear) give identical results up to roundoff errors.

We solve \eqs{daT}{duT} not
by setting $\meanBB$ to the actual mean field resulting from the solutions of
\eqs{dAA}{dUU}, but by setting it to one of four test fields, $\BB^{\rm T}$.
Those are
\EQA
&\BB^{(1)}=(\cos k_B z,0,0),\quad
&\BB^{(2)}=(\sin k_B z,0,0),\\
&\BB^{(3)}=(0,\cos k_B z,0),\quad
&\BB^{(4)}=(0,\sin k_B z,0),
\ENA
where $k_B$ is the wavenumber of the test field, being a multiple of $2\pi/L_z$.
From the solutions of \eqs{daT}{duT}
 we can construct the mean electromotive force,
$\meanEMF=\overline{\uu\times\bb}$ and the mean ponderomotive force,
$\meanFFFF=\overline{\jj\times\bb/\rho - \uu\cdot\nab \uu}$, which are then expressed in
terms of the mean field by the ansatzes
\EQA
&\meanemf_i=\alpha_{ij}\meanB_j-\eta_{ij}\meanJ_j,\label{alpeta}\\
&\meanFFF_i=\phi_{ij}\meanB_j-\psi_{ij}\meanJ_j,\label{phipsi}
\ENA
where $i,j$ adopt only the values $1,2$ as a consequence of setting $\meanB_z$, which is constant anyway, arbitrarily to zero.
Hence, each of the four tensors, $\alpha_{ij}$, $\eta_{ij}$, $\phi_{ij}$,
$\psi_{ij}$, has four components, i.e., altogether we have 16 unknowns.
Note that often the $\alpha$ and $\eta$ tensors are defined as just the symmetric parts of our $\alpha_{ij}$ and $\eta_{ij}$
while their antisymmetric parts are cast into the vectorial coefficients of the $\gamma$ and $\delta$ effects. 
The coefficients $\boldsymbol{\alpha}$, $\boldsymbol{\beta}$, $\boldsymbol{\gamma}$, and $\boldsymbol{\delta}$
describe in turn the effects of turbulent generation, diffusion, pumping and the (non-generative, non-dissipative) so-called R\"adler effect.
In the presence of shear, the coefficient $\eta_{yx}$ plays a prominent role; see \Sec{sect:dynamo}.
In spite of what could be expected from the Lorentz force, being quadratic in $\BB$, the turbulent ponderomotive force \eq{phipsi}
is linear in $\meanBB$. This is because of the presence of the magnetic background turbulence $\bb_0$ via, in the kinematic limit, $\overline{\jj_0\times\bb_{\bm{\bar{B}}} + \jj_{\bm{\bar{B}}}\times\bb_0}$.
}

\subsubsection{Quasikinematic TFM}\label{sec:qktfm}

We state here for comparison the governing equations for the {\QKTFM}
\cite[see also][]{Schrinner05,Schrinner07}.
They consist of just \eq{daT} with $\ffM=\bm{0}$,
but not \eq{duT}, and \eq{alpeta}.
Then, \eq{meanEMFBbar} reduces simply to
\EQ
\meanEMF_{\bm{\bar{B}}}=\overline{\uu\times\bb_{\bm{\bar{B}}}} ,
\EN
and we find the contribution $\overline{\uu_{\bm{\bar{B}}}\times\bb_0}$ missing.
Again, for further details see RB10.

\subsubsection{Resetting}
\label{Resetting}

The test problems \eqs{daT}{duT} are often unstable, but this does not necessarily affect the values
of the resulting turbulent transport coefficients:
They usually show statistically stationary
behavior over limited time spans although the test 
solutions are already growing.
For safety reasons, we always reset them to zero in regular intervals
(typically every 0.5 viscous times); see \cite{Hub+09} for a discussion.
To mask the initial transient, we also remove 20\% of the data from the beginning of each resetting interval.

\subsection{Forcing} \label{forcing}

The standard forcing, implemented in the {\sc Pencil Code}, employs 
white-in-time ``frozen" harmonic plane waves, here restricted to be non-helical.
Their wavevectors are randomly selected from a
thin shell in $k$ space
of radius $\kf$ such that they fit into the periodic computational domain
\citep[for details see, e.g.][]{Kapylaetal20}.
In most of our simulations, we apply this recipe for both
kinetic and magnetic forcings in Equations \eqref{dAA}, \eqref{dAA}, \eqref{daT}, and \eqref{duT}.
The wavevectors are further selected such that no mean field or mean flow is directly
sustained, that is, the case  $k_y=0$ is excluded.\footnote{Without shear, only those with $k_x=k_y=0$
had to be excluded, but due to shear-periodicity,
$2\pi/k_x$ is no longer an integer fraction of $L_x$.}
However, due to roundoff errors, it is unavoidable that averages over harmonic functions
deviate slightly from zero. 
We call this effect ``leakage of the forcing into the mean fields".
Strong shear could produce a linearly growing $\meanB_y$ out of a small $\meanB_x$ due to such leakage.
This is why we checked  
its effect in purely magnetic runs 
and found the growth of $\meanB_y$ to be limited and both components to stay within
margins close to numerical precision.
Nevertheless, as will be discussed in Sect.~\ref{overall},
with this magnetic forcing setup,
the mean magnetic fields very quickly 
(in a few turnover times) reach
dynamically effective strengths
without showing a clear exponential stage.

Hence, another forcing setup was designed, referred to as ``decimated forcing".
In addition to ensuring that $k_y=0$ is excluded, we took out 
all those wavevectors for which $|k_{x,y,z}| \leq k_{\rm min}=2 k_1$.
As will be discussed in the results section, the decimated
forcing has the advantage of reducing the amplitude of the mean fields
generated during the initial stages, thus allowing us to determine
the growth rate of an exponentially growing dynamo instability.
While the standard choice is expected to provide a good approximation to homogeneous isotropic 
velocity turbulence, isotropy could be lost in the decimated case,
given that all wavevectors are parallel or almost
parallel to the spatial diagonals of the box.

However, as 
is discussed in Appendix~\ref{sect:forcings},
the generated turbulence does not markedly deviate
from that by the standard forcing in terms of isotropy.
Also, repeating the 
kinetically forced runs 
($\bm{f}_{\rm M}=\FF_{\rm M} = \bm 0$)   
with decimated forcing does not significantly alter the 
dynamo solutions.

\subsection{Mean flow removal}

\label{removal}
In every case, be it full or simplified MHD, the first instability to be excited is the generation of a mean flow
with horizontal components. These are most likely signatures of the vorticity
dynamo \citep[see, e.g.,][]{Elp03,KMB09}.
As it can destabilize the test problems, we have decided to suppress 
the mean flow by subtracting it from the solution $\UU$ in every time step,
which also avoids leakage of the forcing into $\meanUU$.
With respect to a possible effect on the magnetic field, we refer to
\cite{You08b}, 
who reported, for a very similar simulation setup to that used here,
that the presence of $\meanUU$ did not significantly change the 
properties of the shear dynamo; see their Section~3.4.

\subsection{Input and output quantities}

The simulations are fully defined by choosing the shear parameter $S$,
the forcing setup, amplitude, and
wavenumber, $\kf$, the kinematic viscosity $\nu$,
and the magnetic diffusivity $\eta$.
For normalizations we use the 
horizontal
length scale $k_1^{-1}$, with $k_1=2\pi/L_x$, which is connected
to the vertical length scale  
$k_{1z}^{-1}={\cal A}/k_1$, 
and the viscous time scale $T_\nu=(\nu k_1^2)^{-1}$.
The boundary conditions are (shearing) periodic in all three directions.
The following quantities are used as diagnostics.
We quantify the strength of the turbulence 
by the fluid
and magnetic Reynolds numbers
\begin{equation}
\Rey=\frac{\urms}{\nu \kf},\ \ \ \Rm=\frac{\urms}{\eta \kf}=\Pm \, \Rey,
\end{equation}
where 
\begin{eqnarray}
\Pm=\frac{\nu}{\eta},
\end{eqnarray}
is the magnetic Prandtl number.
The Lundquist number and its ratio 
to
$\Rm$ are given by
\begin{equation}
\Lu=\frac{\Brms}{\eta\kf\sqrt{\rho}}, \quad \frac{\Lu}{\Rm}=\frac{\Brms}{\urms\sqrt{\rho}},
\end{equation}
which is only used in SMHD,
where $\rho=\const$.
The strength of
the imposed shear is measured by the dynamic shear number
\begin{eqnarray}
\ShK=\frac{S}{\urms \kf}.
\end{eqnarray}
As in earlier work, we normalize the turbulent magnetic diffusivity tensor by
the SOCA estimate
\begin{eqnarray}
\eta_0=\urms/3\kf
\end{eqnarray}
or the molecular diffusivity $\eta$.

We define the root-mean-square (rms) value
of the magnetic field as 
$B_{\rm rms} = \langle {\bm B}^2\rangle^{1/2}$ while
$\meanB_{i,{\rm rms}}=\big\langle \meanB_i^2 \big\rangle_z^{1/2}$ are the
rms values of the mean field components.
$\langle.\rangle$ denotes volume averaging and 
$\langle.\rangle_\xi$ 
averaging over a coordinate $\xi$.
The magnetic field is normalized
by the equipartition field strength,
$B_{\rm eq}=\langle \langle \rho\uu^2 \rangle ^{1/2}\rangle_t$.
For the velocity field, we define a time-averaged rms value
$u_{\rm rms} = \langle \langle {\bm u}^2\rangle^{1/2}\rangle_{t}$.
Kinematic dynamo growth rates $\lambda$ are defined as
$\langle d_t \log \Brms \rangle_t$ or $\langle d_t \log \meanB_{i,{\rm rms}} \rangle_t$.

\begin{table*}[t!]\caption{
Summary of the runs with  constant shear and forcing wavenumber.
}
\begin{center}
\begin{tabular}{lrrccrrcc} \hline \hline
Run  &$\Rm$  &$\lambda/(\eta_0 \kf^2)$ & $\eta_{xx}/\eta_0$ & $\eta_{yy}/\eta_0$ & $\eta_{yx}/\eta_0$\phantom{aaa} & $\eta_{xy}/\eta_0$\phantom{aaa} &$\alpha_{\rm rms}/\eta_0 \kf$  &$\eta_{\rm rms}/\eta_0$\\ \hline
\hline
FK1a & 2.1 &$-$0.0354 &0.557$\pm$0.006 &0.547$\pm$0.007 &   0.048$\pm$0.001 &0.351$\pm$0.009 &0.018$\pm$0.009 &0.054$\pm$0.013\\ 
FK1b &11.9 &   0.0140 &0.608$\pm$0.015 &0.598$\pm$0.014 &   0.023$\pm$0.001 &0.419$\pm$0.032 &0.022$\pm$0.011 &0.031$\pm$0.012 \\ 
FK8a & 2.1 &$-$0.0008 &0.572$\pm$0.010 &0.563$\pm$0.011 &   0.044$\pm$0.002 &0.378$\pm$0.009 &0.001$\pm$0.002 &0.048$\pm$0.014 \\ 
FK8b &12.7 &   0.0166 &0.641$\pm$0.019 &0.634$\pm$0.017 &   0.023$\pm$0.001 &0.473$\pm$0.024 &0.009$\pm$0.005 &0.026$\pm$0.009\\ \hline 
SK1a & 2.0 &   0.0006 &0.367$\pm$0.001 &0.393$\pm$0.002 &$-$0.003$\pm$0.000 &0.279$\pm$0.002 &0.021$\pm$0.004 &0.009$\pm$0.001 \\ 
SK1b &12.3 &   0.0183 &0.440$\pm$0.004 &0.412$\pm$0.001 &$-$0.011$\pm$0.002 &0.461$\pm$0.009 &0.020$\pm$0.009 &0.017$\pm$0.009 \\ 
SK4a & 2.1 &$-$0.0042 &0.367$\pm$0.003 &0.390$\pm$0.003 &$-$0.004$\pm$0.000 &0.279$\pm$0.003 &0.008$\pm$0.002 &0.006$\pm$0.001 \\
SK4b &13.3 &   0.0185 &0.334$\pm$0.037 &0.339$\pm$0.044 &$-$0.004$\pm$0.005 &0.239$\pm$0.073 &0.008$\pm$0.004 &0.007$\pm$0.008\\ 
SK8a & 2.1 &   0.0033 &0.367$\pm$0.003 &0.390$\pm$0.004 &$-$0.003$\pm$0.000 &0.274$\pm$0.003 &0.006$\pm$0.002 &0.005$\pm$0.002 \\
SK8b &12.8 &   0.0192 &0.401$\pm$0.005 &0.424$\pm$0.005 &$-$0.015$\pm$0.000 &0.367$\pm$0.010 &0.007$\pm$0.002 &0.017$\pm$0.004 \\  \hline 
SKM1a &1.9 &---\phantom{aa}& 1.794$\pm$0.039 & 1.278$\pm$0.045 & 0.200$\pm$0.025 & $-$0.725$\pm$0.083 & 0.010$\pm$0.055 & 0.250$\pm$0.090 \\ 
SKM4a &2.1 &---\phantom{aa}& 2.012$\pm$0.179 & 1.191$\pm$0.014 & 0.221$\pm$0.012 & $-$0.560$\pm$0.015 & 0.046$\pm$0.017 & 0.230$\pm$0.072 \\ 
SKM8a &1.8 &---\phantom{aa}& 3.054$\pm$0.625 & 1.481$\pm$0.131 & 0.338$\pm$0.064 & $-$0.186$\pm$0.045 & 0.036$\pm$0.011 & 0.352$\pm$0.213 \\ 
SKM16a&2.0 &---\phantom{aa}& 2.238$\pm$0.552 & 1.215$\pm$0.010 & 0.249$\pm$0.062 & $-$0.580$\pm$0.055 & 0.022$\pm$0.008 & 0.260$\pm$0.191 \\ 
\hline 
SKM1ad &2.1 &0.0103 &1.228$\pm$0.214 &1.326$\pm$0.074 &0.247$\pm$0.043 &0.237$\pm$0.117 &0.149$\pm$0.062 &0.441$\pm$0.212 \\ 
SKM4ad &1.9 &0.0315 &1.279$\pm$0.150 &1.455$\pm$0.066 &0.222$\pm$0.022 &0.369$\pm$0.072 &0.081$\pm$0.017 &0.270$\pm$0.119 \\ 
SKM8ad &1.5 &0.0948 &1.688$\pm$0.165 &2.040$\pm$0.150 &0.516$\pm$0.061 &0.383$\pm$0.154 &0.111$\pm$0.069 &0.543$\pm$0.260 \\ 
SKM16ad &1.9 &0.0344 &1.231$\pm$0.070 &1.589$\pm$0.019 & 0.364$\pm$0.116 & 0.279$\pm$0.026 & 0.033$\pm$0.015 & 0.292$\pm$0.015 \\
\hline \hline
\label{models}\end{tabular}
\end{center}
Note: 
For all runs, $\kf/k_1=5$ and $S=-25/T_\nu$
yielding a roughly invariable  
$\ShK$ of $-1.6$. In 
runs with labels ``a", the magnetic Prandtl number 
$\Pm$ is 
$10/3$, 
while for ``b" it is 20.   
The integer in the run name indicates the aspect ratio $\calA$ and the letter ``d" at the end refers to ``decimated forcing".
The number of grid points in the $\calA=1$ models was $144^3$,
while the number of grid points in the vertical ($z$) direction
was increased as  
$\calA$
increased, so that
288, 576, and 1152 grid points in 
$z$
were used for
$\calA=$4,8, and 16, respectively. 
The Mach number in the simulations
was around 0.03 and 0.04 for low and high $\Rm$ runs, respectively.
Due to the low $\Rm$ in all the models investigated, no   
small-scale dynamo 
could get excited, because for flows with low Mach number
the critical $\Rm$ for this instability is around 30 \citep[see, e.g., ][]{HBM04}.
\end{table*}

\section{Results}
\enlargethispage{2\baselineskip}

The naming of the runs is such that the first letter, F or S, indicates
full or simplified MHD, while the second and third refer to the
forcing regime: K and KM referring to purely kinetic
and combined kinetic and magnetic forcing (henceforth magneto-kinetic)
with amplitudes equal up to a factor of $\rho^{-1/2}$ in the magnetic force,
respectively.
The number following the letters indicates the vertical
aspect ratio $\calA$ of the box. 
A trailing letter ``d" stands for ``decimated forcing".
In FMHD, the sound speed was set to $100/k_1 T_\nu$.

\subsection{Overall behavior of the main runs}\label{overall}

As our starting point, we 
defined a setup, related to one
from \cite{SB15b}, 
with marginal dynamo excitation (in incompressible MHD)
and an aspect ratio 
$\calA=8$.
We denote this run as FK8a, and tabulate
 $\Rm$, the growth rate of the initial kinematic stage, $\lambda$, and the $\bm\eta$ components measured by QKTFM
 in Table~\ref{models}.
As reported by \cite{SB15b}, we also observe an initial decay of the rms and mean magnetic
fields, but later we find temporary saturation at very low values,
after which 
a very slow decay is observed, 
indicative of a nearly marginally excited dynamo state. 
Because of the finite $\meanB_x$ present at all times, a much stronger 
(roughly 40 times) $\meanB_y$ is maintained due
to the shear, but as the dynamo is 
nearly marginal, these mean fields 
remain at very low strengths. 
We note that most of the magnetic energy is in the
mean fields, while only a small fraction (less than 20\%) is in the fluctuations.

Next, we repeat this run, but with SMHD, 
which yields Run~SK8a in Table~\ref{models}.
Now rms and mean fields grow, the mean radial and azimuthal components
showing exponential growth at the same rate, albeit still very slow. Nevertheless,
the dynamo instability is somewhat easier to excite than in FMHD.
The azimuthal component is again
much stronger than the radial one with the ratio 
$\meanB_{y, {\rm rms}}$/$\meanB_{x, {\rm rms}}$
similar to the FMHD case.

We continue by repeating these runs with decreased magnetic diffusivity, 
resulting in roughly six times larger magnetic Reynolds number, $\Rm$
(Runs~FK8b and SK8b).
In both simulations we observe exponential growth
of the rms and mean magnetic fields, 
somewhat  faster with SMHD than with FMHD.
The ratio
of the mean field energy to the energy in the fluctuations
remains unchanged with respect to the lower $\Rm$ (and $\Pm$) runs.
We also determine the fastest growing dynamo mode and 
its vertical wavenumber $k_z$
and list them in Table~\ref{dynamonumbers}; 
the fastest growing mode is nearly the same with $k_z/k_{1z}=9$
in both models.
Hence, we can conclude that going from FMHD to SMHD retains the 
dynamo mode, but changes its excitation condition and growth rate somewhat.

As the dynamo growth is slow, simulations with $\calA=8$
are too costly to be run until saturation. 
Hence,
to investigate whether with reduced $\calA$ the dynamo mode could be retained,
 we repeated the runs with
$\calA=1$
(Runs~FK1a, FK1b, SK1a, and SK1b). As is evident
from Tables~\ref{models} and \ref{dynamonumbers}, these runs behave very much like
their tall-box counterparts, the low-$\Rm$ FMHD model being slightly subcritical 
and the high-$\Rm$ one supercritical, while the SMHD runs are both supercritical. 
The fastest growing 
mode now has $k_z/k_1=1$
corresponding to $k_z/k_{1z}=8$ in the tall box. 
We also perform a set of 
runs in SMHD with $\calA=4$;
see Runs~SK4a and SK4b. The former exhibits a very slowly decaying
solution instead of a growing one, which is 
an anomaly in the SMHD set, but
the latter one again exhibits a growth rate very similar to the cubic (SK1b) and 
tall-box (SK8b) cases, 
both with a wavenumber $k_z/k_{1z}=4$. 
All in all, the ``b" runs give rather clear evidence 
that the cubic simulation domains retain the same dynamo mode
as the taller ones.

\begin{figure}\begin{center}
\includegraphics[width=\columnwidth]{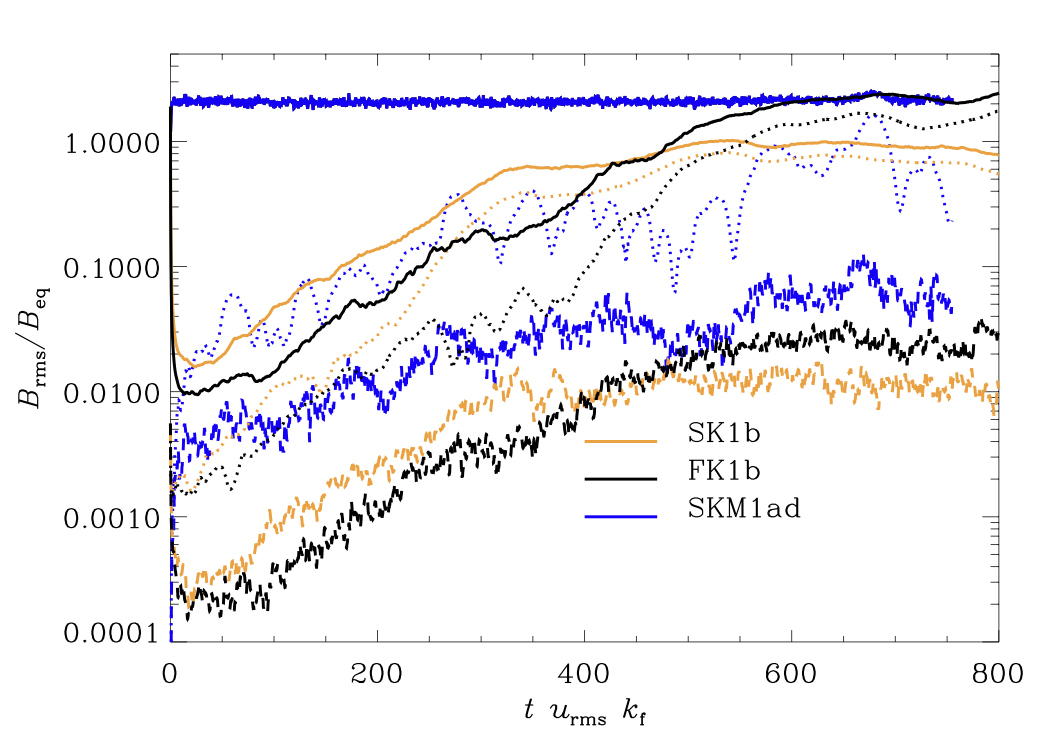}
\includegraphics[width=\columnwidth]{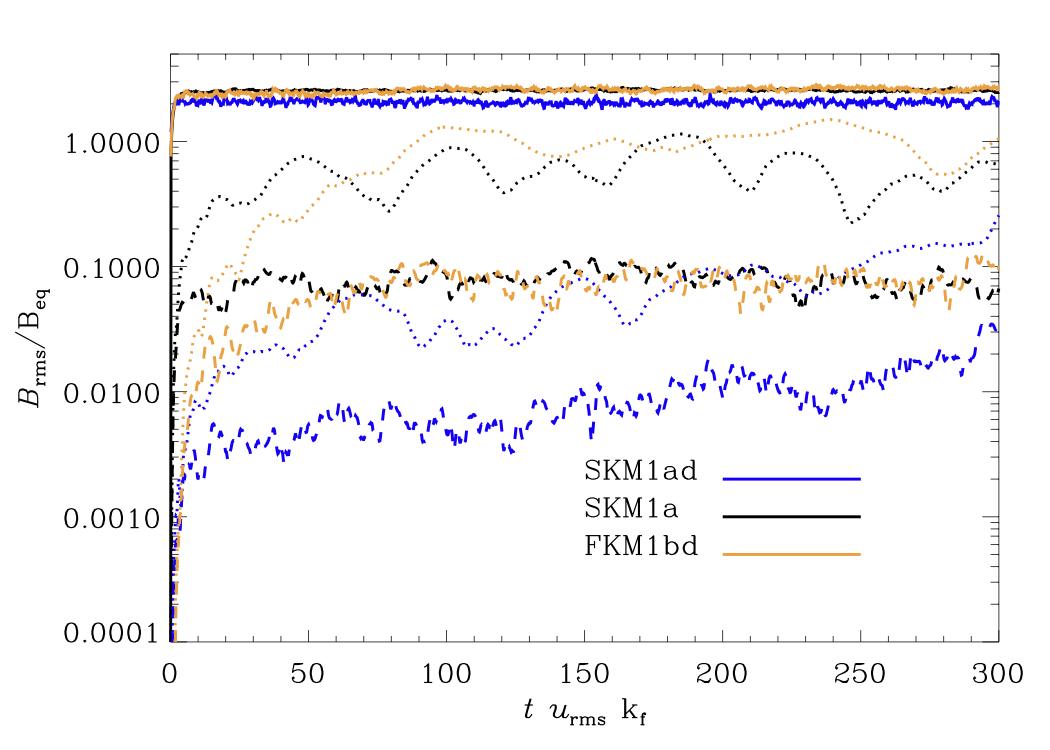}
\end{center}\caption{
Time evolution of rms 
and mean magnetic field strengths from different runs. 
Top: 
comparison of a higher--$\Rm$ FK (black)
and SK (orange) run,
and
a lower-$\Rm$ decimated SKM run (blue).
Bottom:
comparison of SKM runs with $\calA=1$,
with standard (black, Run~SKM1a) and decimated  
(blue, Run~SKM1ad) forcing.
Solid -- $\Brms$,
dotted -- $\meanB_{y, {\rm rms}}$,   
dashed -- $\meanB_{x,{\rm rms}}$.
}\label{fig:Enec}
\vspace{8mm}
\end{figure}

\begin{figure*}\begin{center}
\includegraphics[width=\columnwidth]{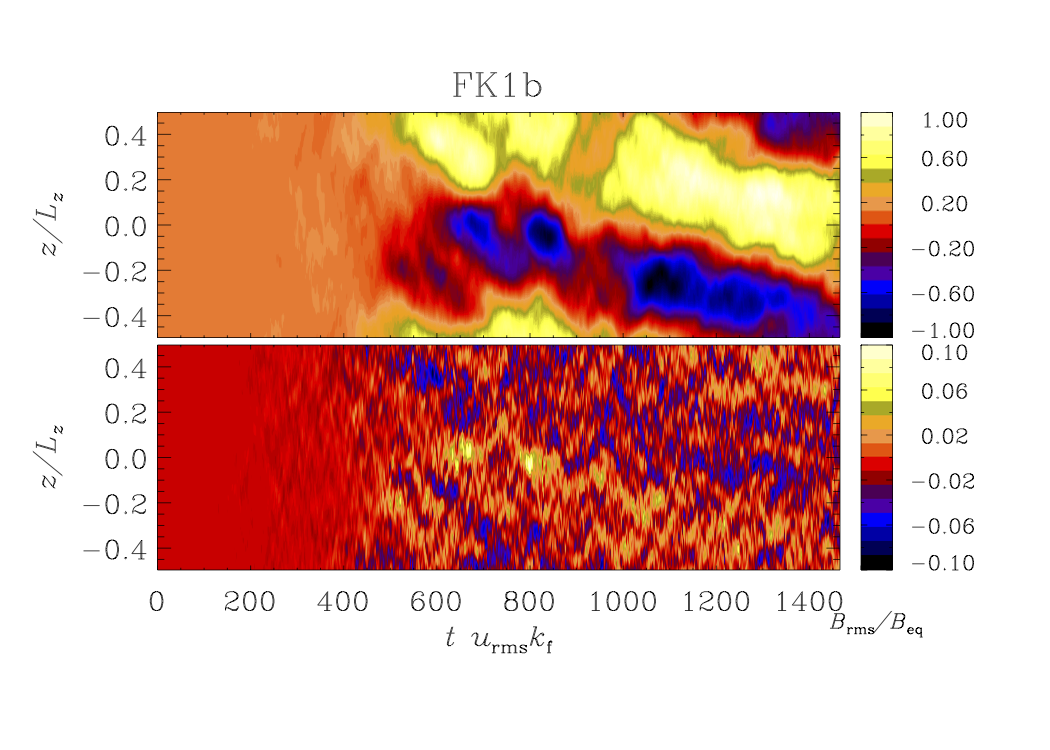}
\includegraphics[width=\columnwidth]{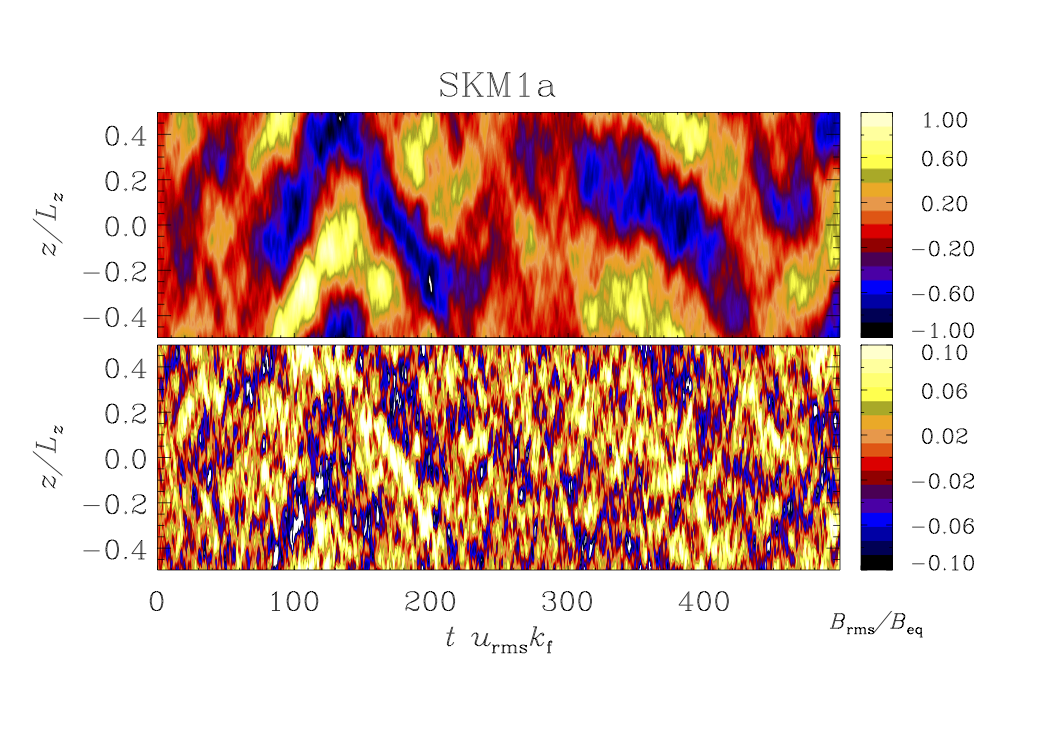}\\
\vspace*{-0.8cm}
\includegraphics[width=\columnwidth]{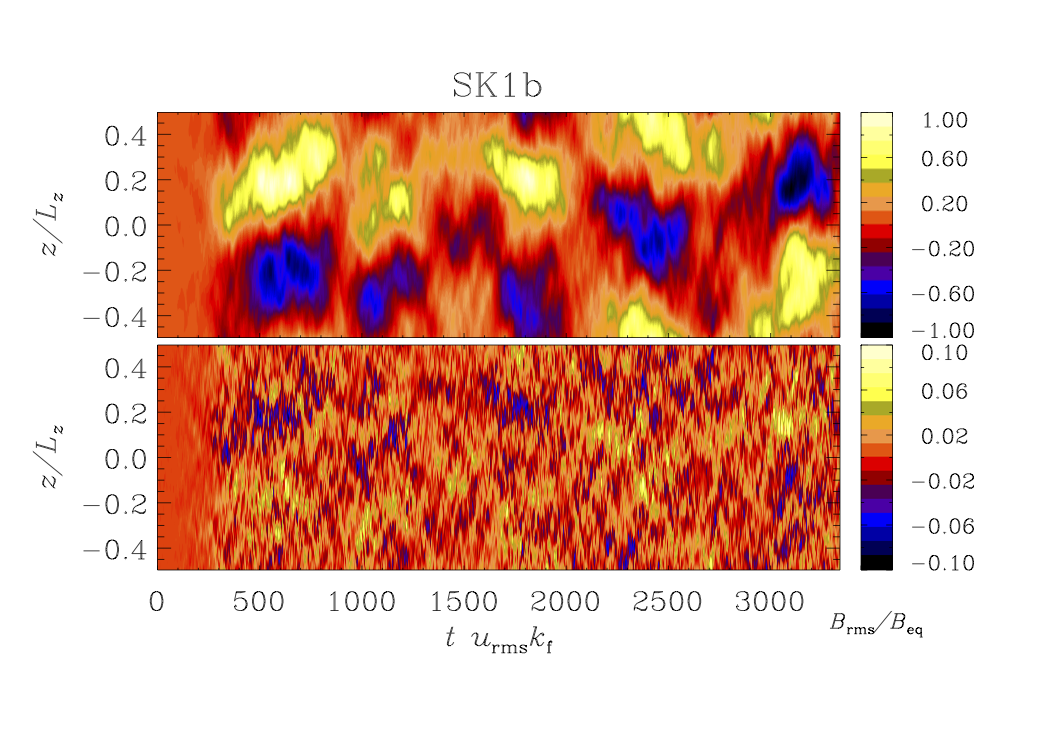}
\includegraphics[width=\columnwidth]{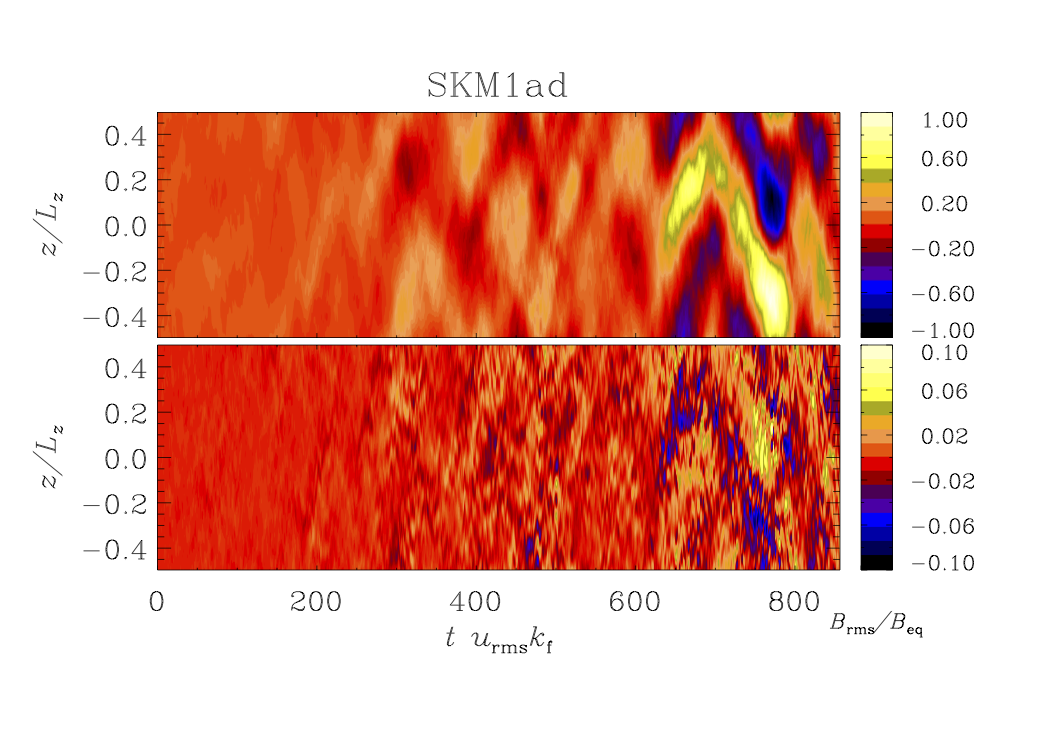}\\
\vspace*{-0.8cm}
\includegraphics[width=\columnwidth]{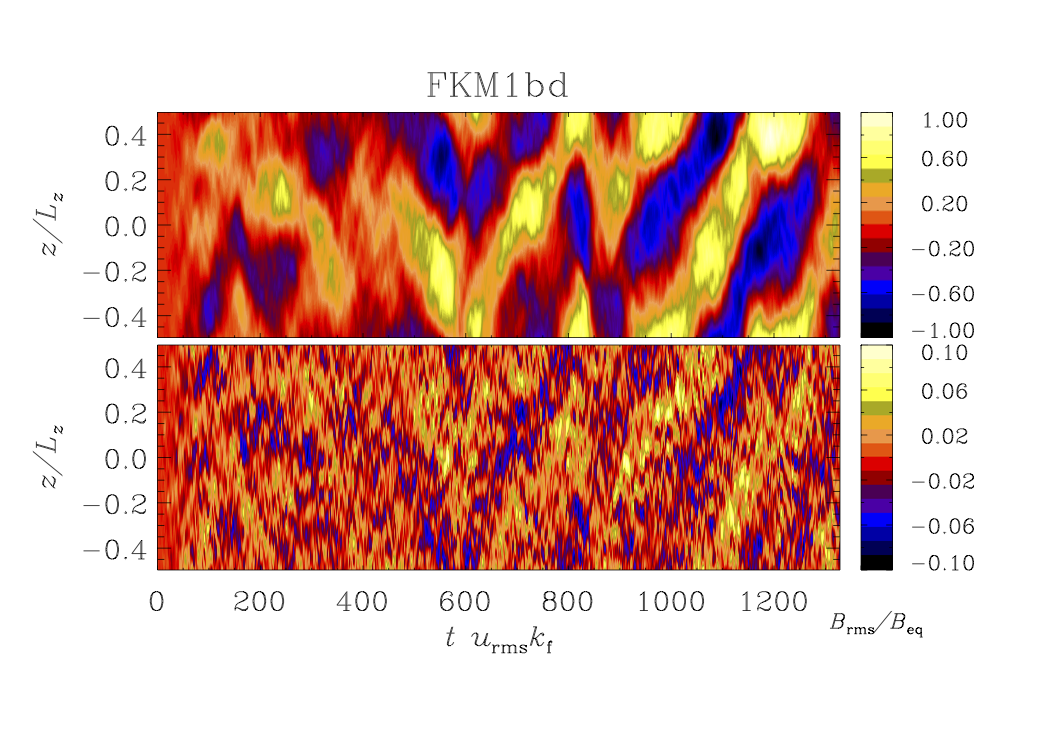}
\vspace*{-0.8cm}
\end{center}
\caption{
Butterfly ($zt$) diagrams of $\meanB_y$ (upper plot in each panel) and $\meanB_x$
(lower plot in each panel). 
Run~FK1b is kinetically forced FMHD, 
Run~SK1b is kinetically forced SMHD, 
SKM1a 
magneto-kinetically forced SMHD, 
SKM1ad its counterpart with decimated forcing, and
FKM1bd is a magneto-kinetically forced run in FMHD, with parameters close
to SKM1ad and FK1b and decimated forcing.
}\label{fig:but}
\vspace{8mm}
\end{figure*}

The time evolution of the rms and mean fields from the cubic runs,
integrated until saturation,
is shown in the top panel of \Fig{fig:Enec} 
with solid and broken lines, respectively.
The growth rate of the SMHD run is somewhat larger, 
but the saturation strength is 
lower than in FMHD. The ratio 
$\meanB_{y, {\rm rms}}$/$\meanB_{x, {\rm rms}}$,
however, is the same.
We also show the mean fields
in a $zt$ diagram in \Fig{fig:but}, top panel. We see the emergence and saturation of the 
Fourier mode with
wavenumber
$k_z/k_1=1$   
in both the radial and azimuthal components, where
each negative (positive) patch of  
$\meanB_y$ is accompanied by a much weaker positive 
(negative) patch in $\meanB_x$.
The patches disappear and reappear quasi-periodically, and also their vertical
position is not constant. 
Our kinetically forced FMHD runs reproduce earlier results 
of similar systems (compare the upper leftmost panel 
of our \Fig{fig:but} to Figure~7 of \cite{BRRK08})
with rather coherent patches in $\meanB_y$, while the SMHD counterpart
(middle left panel of \Fig{fig:but}) 
shows a somewhat more erratic pattern; here, 
however, we must note that the time series 
of the SMHD run is much 
longer. These results are in disagreement with the purely kinetically forced,
incompressible runs of \citet[][their Figure~9(a)]{SB15b}, which show a 
much more erratic pattern than what we observe in either FMHD or SMHD.

Finally, we repeat the simulations,
labeled ``a" ($\Pm=10/3$),  
with the same parameters, but using the magnetic forcing
in addition to the kinetic one, so that 
the same rms velocity is obtained as in the 
kinetically forced cases.
This set of parameters should very closely correspond to the case studied in Figure~9(d) of \cite{SB15b}.
As  seen there, too, we observe a nearly
immediate appearance (during the first five turnover times)
of a strong $\meanB_y$
as is shown for Run~SKM1a in \Fig{fig:Enec}, lower panel.
Although \cite{SB15b} did not show the evolution of 
$\meanB_x$,
our results indicate that 
$\meanB_y$
arises due to the
action of the strong shear on  
$\meanB_x$.
After the initial rapid growth, we do not see any further increase of  $\meanB_x$
while linear growth up to $t \urms \kf \approx 170$
and quasi-regular oscillations occur in $\meanB_y$.
Hence, we are not able to report
a growth rate for Run~SKM1a in Table~\ref{models}, and also not for the larger-$\calA$
runs SKM4a, SKM8a and SKM16a for the same reason.
We note that now the energy in the magnetic fluctuations is dominating over the
energy in the mean field, with roughly 70\% of the total contained in the former.

From \Fig{fig:but}, middle panel, we see that, again, the 
$k_z/k_1=1$   
vertical Fourier mode is the 
preferentially excited one, although the patterns seen in the $zt$ plots
are much more short-lived and erratic in time than in the kinetically forced
counterpart SK1a (same figure, top panel).
Remarkably, there is no kinematic stage, but the large-scale pattern
appears nearly instantly.
(Note that the whole time range shown for Run~SKM1a is roughly as long as
the kinematic range exhibited by Run~SK1b.)
The appearance and evolution of 
$\meanB_y$
also disagree
with the results of \cite{SB15b}, who observed 
a much 
less
erratic
pattern to arise 
in a closely matching parameter regime --- see their 
Figure~9(d).

The rapidly emerging mean fields
in the magnetically and kinetically forced runs are  related to the standard
forcing scheme used in all the simulations presented so far.
Even if this scenario could be regarded as a genuine dynamo instability, 
its investigation is out of the
scope of our current numerical setup, because obviously much higher cadence 
in time should  be used in an attempt to follow the possible kinematic stage. 
Also, the simulations should 
be started from a fully matured 
turbulent MHD background
state, because currently the mean-field growth
occurs during the initial transient state, where even turbulence itself is not yet saturated.  

Hence, instead of fully dwelling on the cause of the rapid initial growth, we 
turn to using the decimated forcing with $k_{\min}/k_1=2$, 
and repeat
Run~SKM1a as a decimated version, now denoted SKM1ad and shown in the lower panel of
\Fig{fig:Enec}  (blue lines). We still see the rapid
appearance of the mean fields, but their magnitudes
are now much lower than in the case of our standard forcing, plotted
with black lines in the same figure for comparison. After the
rapid excitation phase, we observe a slow exponential growth of both  
$\meanB_x$ and $\meanB_y$,
reminiscent of the dynamo
instability seen in the 
FK and SK runs.
The growth rate is now larger
than in the kinetically forced counterparts FK1a and SK1a --- see
Table~\ref{models}; when compared with the higher-$\Rm$ runs 
FK1b and SK1b, as can be seen from the upper panel of \Fig{fig:Enec}, the growth rates 
are nearly equal.
We also produced three
more runs with varying aspect ratio $\calA$ 
(SKM1ad, SKM8ad, and SKM16ad), and
notice that the growth rate is increasing with $\calA$
up to 8, but then decreases again.
We further performed a magneto-kinetically forced FMHD run,
where rapidly emergent mean fields are seen in spite of 
using the
decimated forcing (see the bottom panel of Fig.~\ref{fig:but}, 
showing Run~FKM1bd, with parameters corresponding roughly
to Runs~FK1b and SKM1ad).
The emerging 
large-scale field structures
are very similar to those in
the magneto-kinetically forced SMHD cases, 
but
less coherent than in the kinetically forced FMHD case.
Similar to the SMHD cases with 
standard forcing,
the growth rates
are difficult to estimate, but we do note that the dynamo is now easier
to excite than in the kinetically forced FMHD case, where the large-scale field 
emerged 
only
at 600 turnover times instead of a few tens.
Hence, we cannot confirm the finding of \cite{SB15b} that more
coherent structures emerge when one goes from kinetic to magnetic forcing,
as was the case in their incompressible study.

Based on these runs with different forcings, we propose
that the slow dynamo instability could have been
drowned by the stronger initial mean fields when forced 
with the standard forcing.
Although the growth rate of the dynamo instability is similar
to the kinetically forced cases, and the 
wavenumber of
the dynamo instability are the same in both cases, the change of
the growth rate as function of the aspect ratio of the box indicates
that some key properties of the dynamo instability do change when
magnetic forcing is used. In the next section we make an attempt
to investigate what exactly has changed by measuring the turbulent
transport coefficients in the systems with the relevant TFM variant.

 \begin{figure*}\begin{center}
\includegraphics[width=0.45\textwidth]{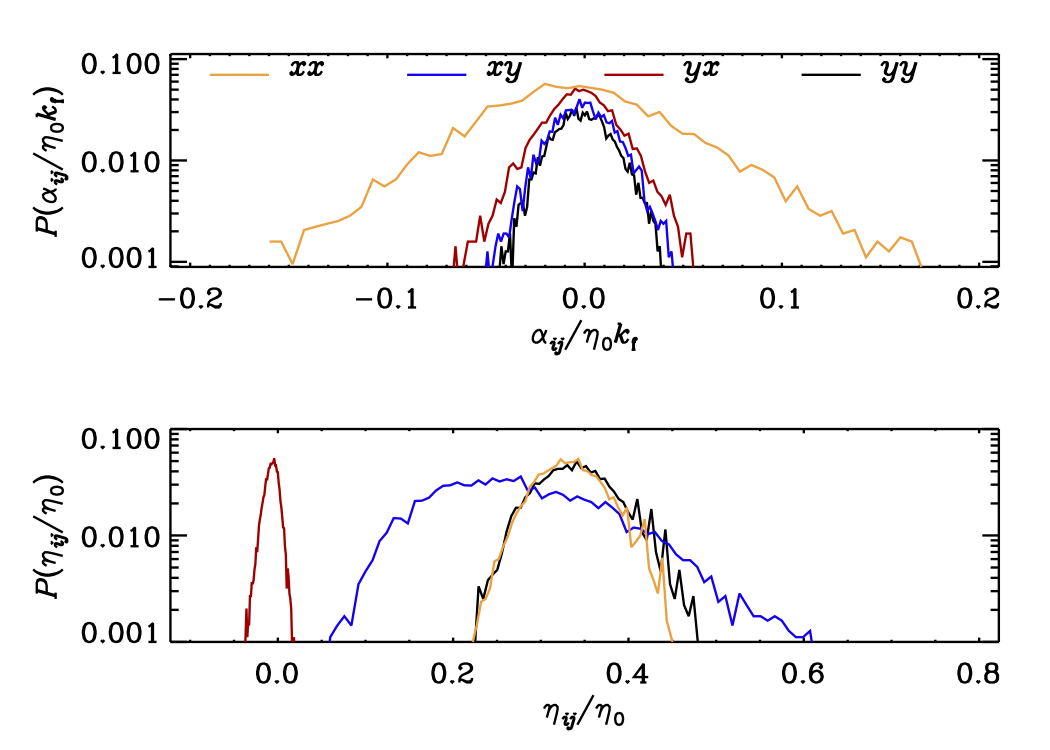}
\includegraphics[width=0.45\textwidth]{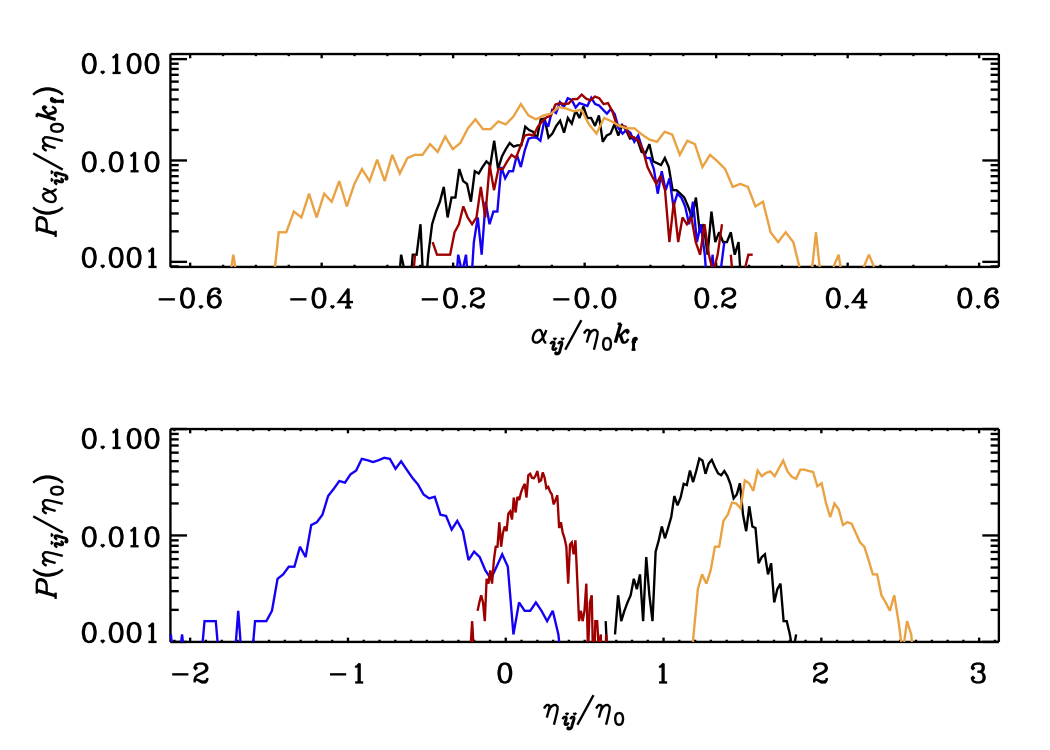}
\end{center}\caption{
Probability density functions of all turbulent transport coefficients. 
Top: $\alpha_{ij}$, bottom: $\eta_{ij}$.
Left: 
kinetically forced SMHD Run~SK1b, right: 
magneto-kinetically forced SMHD Run~SKM1a.
}\label{fig:histo}
\vspace{8mm}
\end{figure*}

\subsection{Turbulent transport coefficients}

\begin{figure*}\begin{center}
\includegraphics[width=0.9\columnwidth]{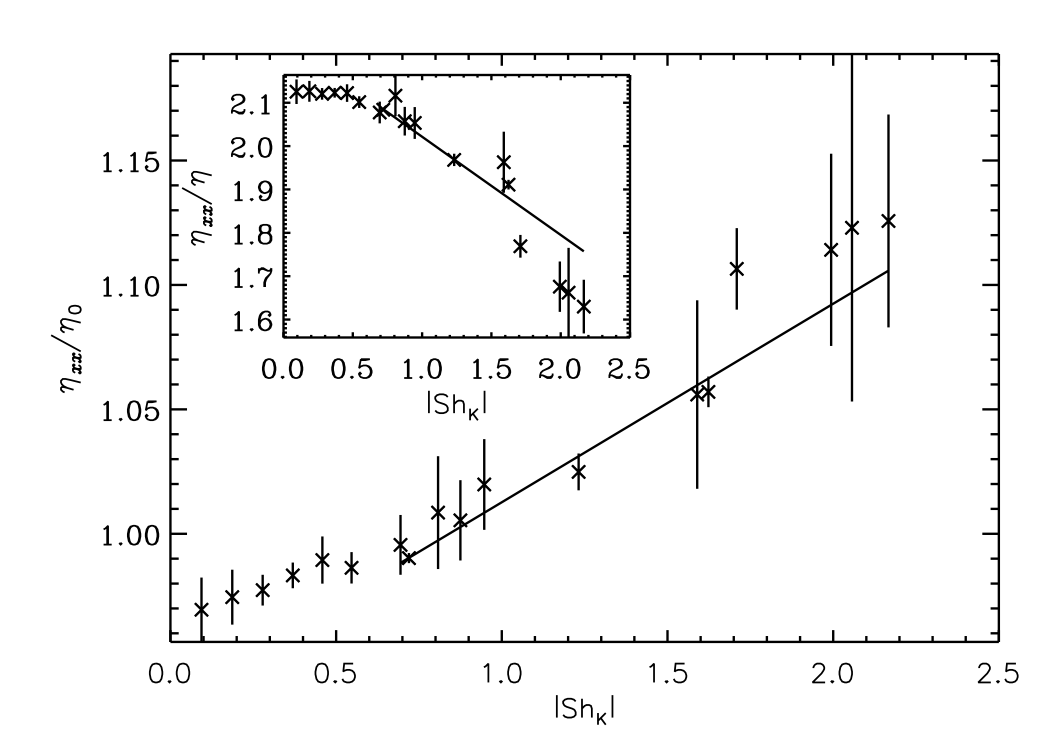}
\includegraphics[width=0.9\columnwidth]{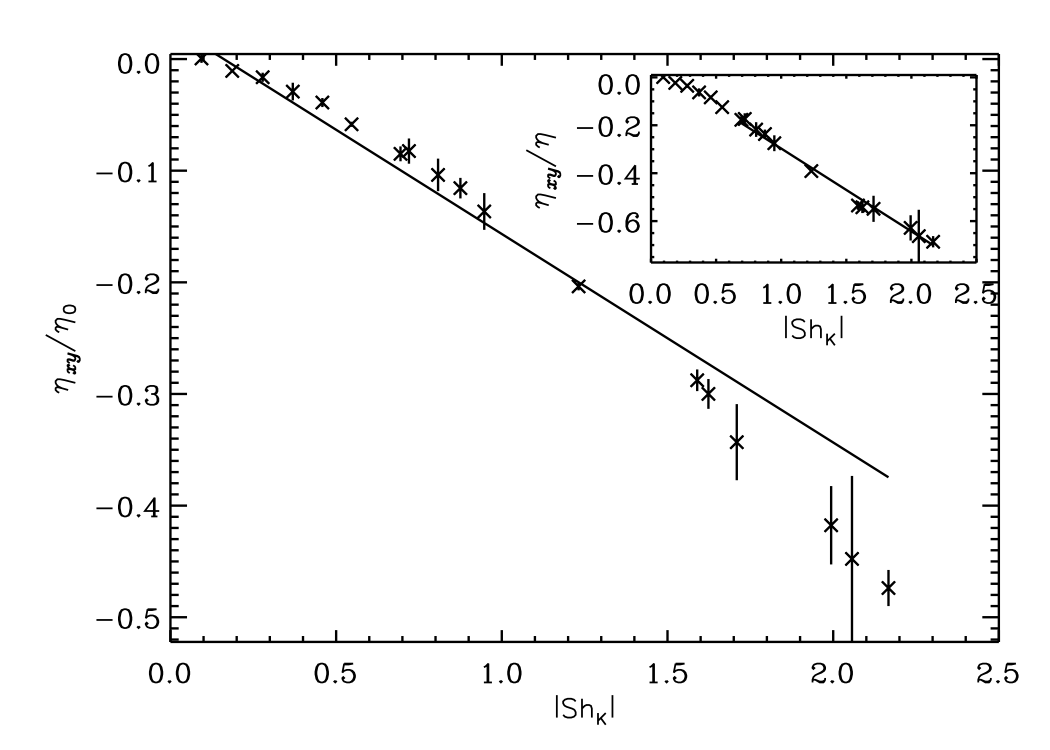}\\
\includegraphics[width=0.9\columnwidth]{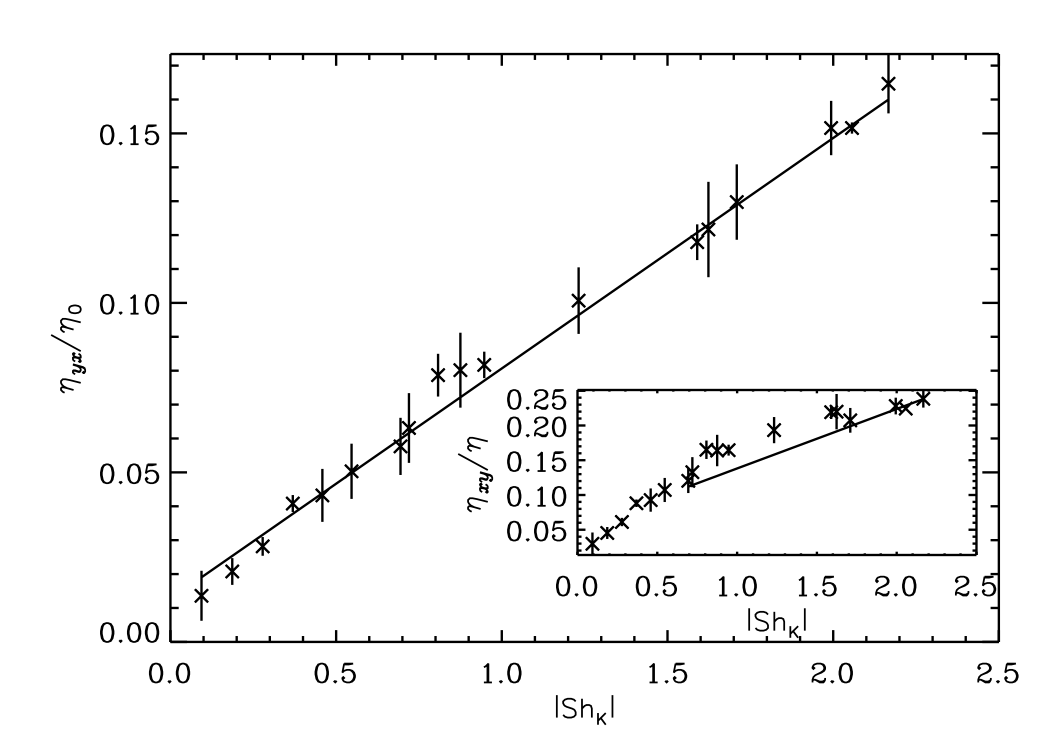}
\includegraphics[width=0.9\columnwidth]{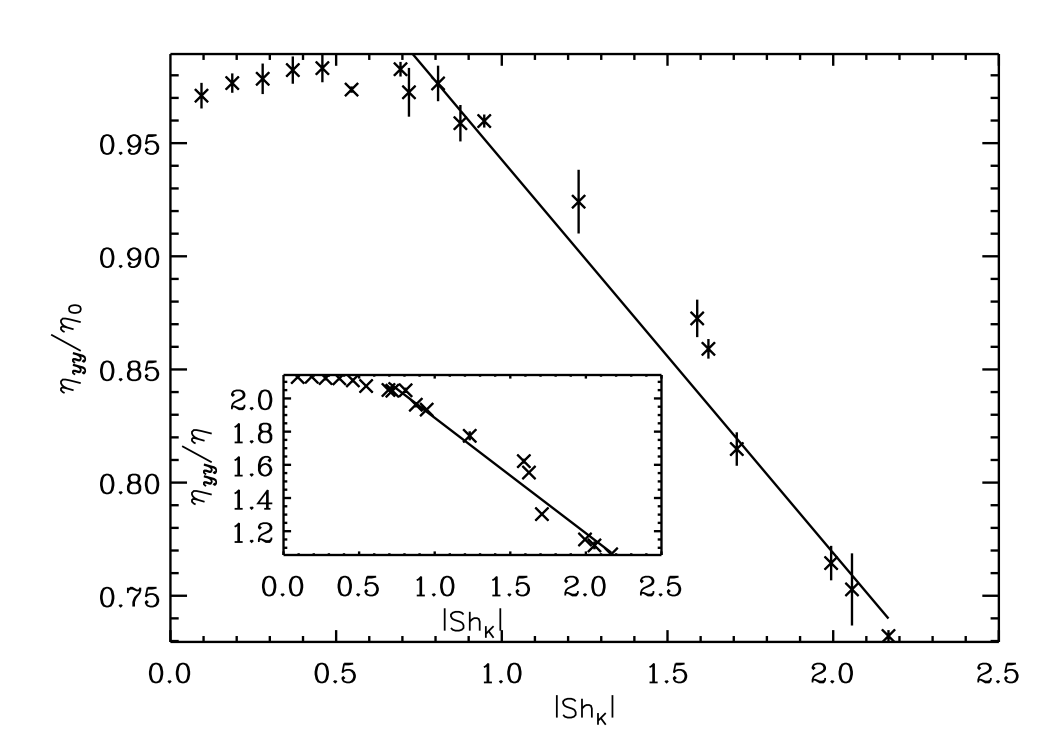}
\end{center}
\caption{
Dependence of the turbulent diffusivity
tensor components, measured with NLTFM,
on the shear number in the magneto-kinetically forced cases.
In the big plots we normalize to the
SOCA estimate $\eta_0$, 
while in the insets to the molecular diffusivity $\eta$.
}\label{fig:eta_S}
\vspace{8mm}
\end{figure*}

  \begin{table*}[t!]\caption{
Summary of the runs with varying shear.}
\begin{tabular}{lccccccc} \hline \hline
Run  &$\ShK$ & $\eta_{xx}/\eta_0$ & $\eta_{yy}/\eta_0$ & $\eta_{yx}/\eta_0$ & $\eta_{xy}/\eta_0$ &$\alpha_{\rm rms}/(\eta_0 \kf)$  &$\eta_{\rm rms}/\eta_0$\\ \hline
\hline 
SKM1a001 & $-$0.094	&2.125$\pm$0.028 	&2.129$\pm$0.012 	&0.030$\pm$0.016 	&\phantom{$-$}0.001$\pm$0.009  &0.094$\pm$0.015 	&0.137$\pm$0.050 \\ 
SKM1a002 & $-$0.187	  &2.126$\pm$0.024 	&2.131$\pm$0.009 	&0.045$\pm$0.009 	&$-$0.023$\pm$0.003 &0.101$\pm$0.029  &0.142$\pm$0.066 \\ 
SKM1a003 & $-$0.278	&2.120$\pm$0.023 	&2.123$\pm$0.015 	&0.061$\pm$0.006 	&$-$0.035$\pm$0.009 &0.092$\pm$0.034 	&0.137$\pm$0.036 \\ 
SKM1a004 & $-$0.369	&2.123$\pm$0.011 	&2.121$\pm$0.013 	&0.088$\pm$0.005 	&$-$0.063$\pm$0.017 &0.096$\pm$0.014 	&0.153$\pm$0.043 \\ 
SKM1a005 & $-$0.458 	&2.122$\pm$0.020 	&2.109$\pm$0.013 	&0.093$\pm$0.017 	&$-$0.084$\pm$0.009 &0.081$\pm$0.033 	&0.147$\pm$0.047 \\ 
SKM1a006 & $-$0.547 	&2.101$\pm$0.013 	&2.074$\pm$0.003 	&0.107$\pm$0.017 	&$-$0.125$\pm$0.005 &0.088$\pm$0.032 	&0.164$\pm$0.071 \\ 
SKM1a007 & $-$0.649      &2.077$\pm$0.025   &2.051$\pm$0.007   &0.120$\pm$0.018
&$-$0.177$\pm$0.024 &0.091$\pm$0.027         &0.173$\pm$0.068 \\
SKM1a008 & $-$0.719 	&2.084$\pm$0.004 	&2.046$\pm$0.023 	&0.133$\pm$0.022 	&$-$0.173$\pm$0.014 &0.084$\pm$0.019 	&0.173$\pm$0.081 \\ 
SKM1a009 & $-$0.808 	&2.116$\pm$0.048 	&2.049$\pm$0.016 	&0.165$\pm$0.013 	&$-$0.218$\pm$0.024 &0.077$\pm$0.032 	&0.196$\pm$0.074 \\ 
SKM1a01  & $-$0.873 	&2.057$\pm$0.033 	&1.962$\pm$0.016 	&0.164$\pm$0.023 	&$-$0.237$\pm$0.031 &0.081$\pm$0.034 	&0.196$\pm$0.083 \\ 
SKM1a011 & $-$0.947 	&2.053$\pm$0.037 	&1.932$\pm$0.006 	&0.165$\pm$0.008 	&$-$0.275$\pm$0.019 &0.080$\pm$0.031 	&0.197$\pm$0.073 \\ 
SKM1a015  & $-$1.226	&1.968$\pm$0.014 	&1.775$\pm$0.027 	&0.193$\pm$0.019 	&$-$0.391$\pm$0.033 &0.074$\pm$0.027 	&0.219$\pm$0.094 \\ 
SKM1a02  & $-$1.582	&1.963$\pm$0.070 	&1.622$\pm$0.015 	&0.219$\pm$0.010 	&$-$0.535$\pm$0.007 &0.067$\pm$0.018 	&0.233$\pm$0.068 \\ 
SKM1a021  & $-$1.623	 &1.911$\pm$0.011 	&1.553$\pm$0.008 	&0.220$\pm$0.025 	&$-$0.542$\pm$0.018 &0.064$\pm$0.030 	&0.238$\pm$0.103 \\ 
SKM1a025  & $-$1.709	 &1.769$\pm$0.026 	&1.303$\pm$0.012 	&0.207$\pm$0.018 	&$-$0.549$\pm$0.055 &0.058$\pm$0.030	&0.223$\pm$0.083 \\ 
SKM1a031  & $-$1.985	&1.676$\pm$0.058 	&1.150$\pm$0.011 	&0.228$\pm$0.012 	&$-$0.628$\pm$0.053 &0.056$\pm$0.024	&0.238$\pm$0.069 \\ 
SKM1a0325 & $-$2.057 &1.662$\pm$0.103 	&1.114$\pm$0.024 	&0.224$\pm$0.002 	&$-$0.663$\pm$0.110 &0.050$\pm$0.014	&0.236$\pm$0.035 \\ 
SKM1a035  & $-$2.156	 &1.630$\pm$0.062 	&1.060$\pm$0.004 	&0.238$\pm$0.013 	&$-$0.686$\pm$0.023 &0.052$\pm$0.012	&0.248$\pm$0.083 \\ 
\hline 
\hline
\label{Shmodels}\end{tabular}
\\
Note:
Forcing wavenumber $\kf/k_1=5$.
The magnetic Reynolds number, $\Rm$, varies from 1.4 (for weak shear) to 2.1 (for strong shear),
and the Lundquist number, $\Lu$, from 4.2 (for weak shear) to 4.8 (for strong shear). $\calA=1$ in all the runs.
\end{table*}

\subsubsection{Cases of strong shear} 

In this subsection we compare cases of strong shear in kinetically forced FMHD and SMHD, and
magneto-kinetically forced SMHD, measured with the appropriate variant of the TFM. 
We choose $S=-25/T_\nu$, 
which, with the selected amplitude of the forcing,
results in the shear number $\ShK \approx -1.6$, indicating a strong influence of 
shear on the system. This setup closely matches the cases investigated by \cite{SB15b}.

First we use the QKTFM to measure the turbulent transport coefficients in the
kinetically forced FMHD cases, the results being presented in Table~\ref{models}.
We measure zero mean in all $\alpha$ 
components, hence we tabulate only
the rms values of the $\alpha$ fluctuations:
$\alpha_{\rm rms} = \langle \alpha_{ij}^2 \rangle_t^{1/2}$.
The same applies to all other runs studied here.
In the low-$\Rm$ cases (FK1a, FK8a) we measure relatively isotropic diagonal components of the 
$\eta$ tensor, positive and somewhat smaller values of $\eta_{xy}$  and much smaller
positive values of  $\eta_{yx}$. 
In these cases, no indication of LSD instability is seen.

In the high $\Rm$ cases (FK1b and FK8b), the 
diagonal components of $\bm\eta$
have, as expected, higher magnitudes,
showing only mild anisotropy, as in the low-$\Rm$ cases, such that $\eta_{xx}$ 
somewhat exceeds $\eta_{yy}$. 
$\eta_{xy}$ is increased with respect
to the diagonal components, reaching
roughly 75\% of their magnitudes.
$\eta_{yx}$ is still positive, and decreases in magnitude.
In these cases we see LSD action, but with $\eta_{yx}$ being positive it seems unlikely that
the dynamo is of SC-effect origin, in agreement with previous
numerical studies \citep{BRRK08,You08,SJ15}.
They did not consider as large values of the
shear parameter as here, so we can now extend this conclusion to the strong-shear regime.
This is consistent with a series of earlier analytical works that
treated shear non-perturbatively and found no evidence of
SC-assisted LSD \citep{SS09a,SS09b,SS10,SS11}.
We analyze the possible dynamo driving mechanism in more detail in Sect.~\ref{sect:dynamo}.

Next we turn to the kinetically forced
SMHD cases, analyzed with both the QKTFM and NLTFM,
yielding consistent results, as discussed in
Appendix~\ref{sect:comparison}. 
The biggest difference from FMHD is that all $\bm\eta$ components
are systematically smaller
in SMHD, and moreover, $\eta_{yx}$ has
changed sign to negative values, being statistically significant
within errors; see Table~\ref{models}.
Also, the rms $\alpha$ values are 
similar or a bit larger.
In the face of the turbulent
transport coefficients, it seems understandable that 
for the low-$\Rm$ cases the LSD is excited in SMHD but not 
 in FMHD, because the diffusive coefficients are lower, while the inductive ones are larger.  
Also, the sign of $\eta_{yx}$ would now be favorable to enable the SC effect to support an LSD. 
Further, it is noteworthy that the diagonal components of $\bm\eta$ become more notably
anisotropic, but now $\eta_{yy}$ mostly exceeds $\eta_{xx}$.
In \Fig{fig:histo}, we show for Run~SK1b the probability density distributions of all
tensor components. The diagonal $\bm\alpha$  components exhibit larger values than the   
off-diagonal ones, $\alpha_{xx}$ being especially strong.
The off-diagonal components are very similar to each other.
The diagonal $\eta$  components are close to being isotropic. 
$\eta_{yx}$ is fluctuating 
tightly around zero, and exhibits a very small negative mean.
The distribution of $\eta_{xy}$ is broad, but always in the positive.
    
Lastly, we turn to the 
magneto-kinetically forced SMHD cases, analyzed with the NLTFM.
In the low-$\Rm$ runs, all  components of  $\bm\eta$ 
show larger magnitudes
than in the kinetically forced cases. 
Its diagonal components 
now show very strong anisotropy, with $\eta_{xx}$ being again dominant over $\eta_{yy}$ as in the FMHD cases. 
$\eta_{xy}$ has changed sign to negative values, while $\eta_{yx}$ is again positive.
The rms values of $\alpha$ and $\eta$ are (mostly) increased, in particular those of the latter. 
The probability density functions of the 
transport coefficients, 
shown in the  right column of \Fig{fig:histo}, 
clearly demonstrate
the
anisotropy of the diagonal components of $\bm\eta$ and
the sign change of $\eta_{xy}$ to large negative values, with $\eta_{yx}$ now exhibiting a 
clearly positive mean 
with some negative values as well. The $\alpha$ components are very similar to the kinetically forced SMHD case, with $\alpha_{xx}$ attaining much larger values than $\alpha_{yy}$ and the off-diagonal components. The positive sign of $\eta_{yx}$ rules out the existence of an SC-effect dynamo
in these cases. As will be discussed in detail in Sect.~\ref{sect:dynamo},
the $\alpha$ and $\eta$ fluctuations 
then remain as possible candidates to provide the necessary ingredients for an LSD.

\subsubsection{Dependence on the shear parameter}

In this section we report on the dependence of the turbulent 
transport coefficients on the shear number $\ShK$
in runs with magneto-kinetical
forcing.
We list our runs, their basic diagnostics, and the 
turbulent transport coefficients measured
with the NLTFM in Table~\ref{Shmodels}.
As the standard forcing was used here, we did not see
any exponential growth in the evolution of the mean fields; see
Sect.~\ref{overall} for a reasoning.
Hence, no growth rates are reported, 
and we note that all the transport coefficients are measured from
a stage where the mean magnetic fields are dynamically significant.
As can be seen from the listed $\Lu$, these runs are all strongly
magnetically dominated, likely because 
the small-scale magnetic fields are primarily generated by the magnetic forcing.
Our purpose is to scan a wider range of shear strengths
 for possible occurrences of a negative $\eta_{yx}$ as a function of 
$\ShK$, which could enable an
SC-driven LSD. The results 
are depicted in \Fig{fig:eta_S}, where we present the 
$\bm\eta$ components in two different normalizations.
As can be seen, with weak shear ($|\ShK| < 0.5$),
the diagonal components of $\bm\eta$ 
are isotropic, while with stronger shear, 
anisotropy develops such that in the SOCA normalization, $\eta_{xx}$ 
increases linearly while $\eta_{yy}$  decreases linearly. 
Normalizing to molecular diffusivity,
both components are decreasing linearly, $\eta_{xx}$
less steeply than $\eta_{yy}$.
For weak shear, $\eta_{yx}$ adopts
small positive values, which keep increasing linearly with
shear in the SOCA normalization. The linear trend is less
clear in the molecular diffusivity normalization.
Furthermore, $\eta_{xy}$ attains weakly negative values for weak shear, and increasingly
negative ones for strong shear. The trend is very close to linear
when molecular diffusivity is used for normalization. 
Hence, we find no possibility for 
an MSC-effect-driven dynamo at any shear number investigated.

The dependences of $\eta_{ij}$ on shear, as obtained here, are in broad
agreement with the results of \citet{SS11} based on an analytical study 
in which arbitrarily large values of the shear parameter $S$
could be explored;
see references therein for more discussion.
The two off-diagonal components $\eta_{xy}$ and $\eta_{yx}$ were found to
start from zero at zero shear and, while the more relevant 
$\eta_{yx}$ increases with $|S|$ to remain positive, 
$\eta_{xy}$ behaves in a more complicated manner than found here,
exhibiting both signs depending on the value of $S$:
It decreases with increasing $|S|$ to become negative
up to a certain value of shear,
as in the present work; we refer the reader to \cite{SS11} for
more detail on its behavior at larger $\left| S \right|$.

\subsubsection{Dependence on the aspect ratio}

We have studied the dependence of the turbulent transport coefficients on the aspect
ratio $\calA$ of the domain in 
the three different cases (FMHD, SMHD with 
kinetic or magneto-kinetic forcing)
with fixed shear parameter $S=-25 /T_\nu$. 
The measured growth rate 
of the rms magnetic field,
which coincides with those of 
of $\meanB_x$ and $\meanB_y$
except for standard magnetic forcing,
and the measured turbulent transport coefficients are listed
in Table~\ref{models};
see runs with labels 4, 8, and 16, indicating $\calA$. 

In the kinetically forced FMHD and SMHD cases, the growth rate of the
magnetic field is largely independent of the aspect ratio of the box,
indicating that always
one and the same dynamo mode is growing.
We also measure the vertical wavenumber $k_z$ of the fastest growing
dynamo mode in the kinematic stage
(see Table~\ref{dynamonumbers}), which
supports this conclusion, as we see the
$k_z/k_{1z}$ increasing
proportional to $\calA$.
The turbulent transport coefficients do not show a marked dependence on 
$\calA$ either. 

\begin{figure}\begin{center}
\includegraphics[width=0.9\columnwidth]{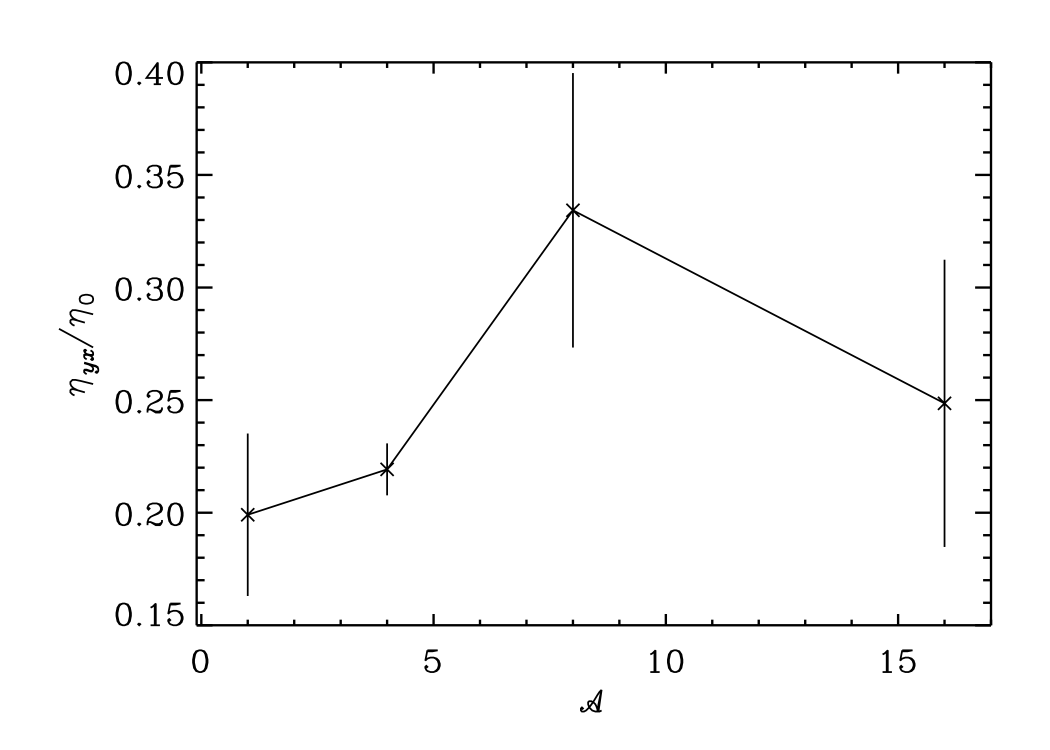}
\end{center}\caption{
Dependence of 
$\eta_{yx}$
on the aspect ratio $\calA$ for SMHD
cases with standard magneto-kinetic forcing.
}\label{fig:eta_aspect}
\vspace{8mm}
\end{figure}

 In  SMHD with standard magneto-kinetic forcing,
the situation is somewhat different. 
As we cannot draw conclusions on the growth rate of the magnetic field in
these cases, we use the corresponding
cases with decimated forcing as a guideline. The latter
(see Table~\ref{models}, runs with label ending in ``d") 
show that the growth rate is increasing 
with $\calA$ up to 8 and then decreases again in the tallest box.
In \Fig{fig:eta_aspect}
we show $\eta_{yx}$ as a function of $\calA$.
It can be seen that the magnitudes of the
turbulent transport coefficients change somewhat
as a function of the aspect ratio, although 
the magnitudes seem to saturate for the tallest box.
The diagonal components grow in magnitude, $\eta_{xx}$ 
somewhat more than  $\eta_{yy}$ 
making the anisotropy in the turbulent diffusivity
even larger. The negative values measured for $\eta_{xy}$ tend to get weaker in taller boxes.
The positive values of $\eta_{yx}$ increase with $\calA$,
hence we see no tendency for larger boxes 
to be more favorable for the SC dynamo.
The fluctuating $\alpha$ and $\eta$
behave similarly, with their magnitudes first increasing, but then decreasing for the tallest box.
The decimated forcing cases show a similar trend for 
$\calA=4, 8,$ and $16$
(SKM4ad, SKM8ad, and SKM16ad) while the case $\calA=1$ (SKM1ad) 
shows higher values of the transport coefficients not agreeing with this trend.

As the number of grid points is proportional to $\calA$ at fixed resolution,
resource limitations dictated to integrate 
the large-$\calA$ runs only over 
significantly shorter time spans.
However, as we have discussed
above, the mean fields grow initially
very rapidly in all  runs with standard forcing,
irrespective of the aspect ratio.
Hence, an effect of the different integration times
on the values of transport coefficients can be ruled out.

One could also speculate that some spatio-temporal nonlocality 
\citep[see, e.g.,][]{RB12} might come into
play with magnetic forcing, but when choosing our forcing wavenumbers, we have
taken care of $\kf$ being scaled with respect to the vertical extent of the computation domain
such that the forcing wavenumber remained constant. Our procedure,
however, does not take into account non-local
effects in any way.

The dependence of the growth rate
on the aspect ratio could also be due to different dynamo modes being excited in 
boxes of different size.
This was found by \cite{Shi16} in a similar context, 
but including rotation,
in which case the turbulence 
was self-sustained (i.e., not
driven as in our study) by the magnetorotational instability (MRI).
\footnote{Note that in
the case of the MRI
 there is no background turbulence,
not even a kinetic one, because the 
whole
turbulence is ``created" by the MRI 
due to the presence of a 
large-scale
field \citep[see, e.g.,][]{BNST95}.
Thus, a 
magnetic SC effect
as defined by \cite{SB15a} 
that has its sole cause in a magnetic background turbulence $\bb_0$,
cannot exist.}
They found the dynamo to be more efficient in taller
boxes, and interpreted this by having ``cut out"  some modes in the smaller boxes. 
However, determining
the vertical wavenumber of the fastest growing mode in the kinematic stage
for the decimated forcing runs, we find no evidence
for this. As the turbulence in the 
cases with standard and decimated forcing is different 
though, we cannot regard this as completely conclusive evidence that
rules out this scenario.

\subsection{Interpretation of the dynamo instability}\label{sect:dynamo}

For SC-effect-driven
dynamos, the dispersion relation from linear stability analysis
for solutions, exponential in time, reads \citep[see, e.g.,][]{BRRK08}
\begin{equation}
\frac{\lambda_{\pm}}{\etaT k_z^2} = -1 \pm \frac{1}{\etaT} \sqrt{\left( \frac{S}{k_z^2} + \eta_{xy}\right) \eta_{yx} + \epsilon^2},
\label{eq:disp}
\end{equation}
with $\etaT = \eta + \etat$, $\etat = \left( \eta_{xx} + \eta_{yy} \right)\!/2$, 
$\epsilon =\left( \eta_{xx}-\eta_{yy} \right)\!/2$. A necessary and sufficient condition
for growing solutions is that the radicand is positive, and larger than $\etaT^2$.
In other words (for $\epsilon\approx 0$)
\begin{equation}
D_{\eta S} \equiv \left(\frac{S}{k_z^2} + \eta_{xy}\right) \frac{\eta_{yx}}{\etaT^2} > 1, \label{Eq:SC}
\end{equation}
which is often further simplified by ignoring the contribution from $\eta_{xy}$, because 
it is considered negligible in comparison to $S/k_z^2$.
This also holds for the systems studied here, but we note that in all our cases, $\eta_{xy}$ is much
larger than $\eta_{yx}$ and in the kinetically forced cases it is even comparable to the diagonal components.
Hence, setting it to zero, as has been done in some fitting experiments 
to determine the turbulent transport coefficients \citep[see, e.g.,][]{Shi16}, is not justified.
Especially in the magneto-kinetically forced cases
with strong shear, the assumption $\epsilon\approx 0$, made in those fitting experiments,
breaks down, too. 

\begin{table}[t!]\caption{
Dynamo numbers for the runs in Table \ref{models}.
 }
\centerline{
\begin{tabular}{llrrr} \hline \hline
Run  & $\!\!\!\!\!\!\!\!{k_z/k_{1z}}$ &$D_{\eta S}$ &$D_{\eta_{\rm rms} S}\!\!\!\!$& $D_{\alpha S}$\\    \hline
FK1a &1*&$-1.4$ &1.6 &2.8 \\ 
FK1b &1 &$-3.9$ &5.3 &19.2\\ 
FK8a &9*&$-1.0$ &1.1 &0.7 \\ 
FK8b &9 &$-2.8$ &3.2 &4.7 \\ \hline 
SK1a  &1 &0.1 &0.3 &2.9\\ 
SK1b &1 &2.7 &4.2 &25.5 \\
SK4a  &4 &0.1 &0.2 &1.5 \\
SK4b &4 &1.3 &2.6 &14.6 \\ 
SK8a &4 &0.5 &0.7 &8.2\\
SK8b &9 &2.1 &2.4 &3.6  \\  \hline 
SKM1a  &1* &$ -2.7$ & 3.3 & 6.8 \\ 
SKM4a  &4* &$ -2.9$ & 3.0 & 3.0 \\ 
SKM8a  &4* &$-12.2$ &12.7 &13.1 \\ 
SKM16a &9* &$ -9.4$ & 9.9 & 7.5 \\ \hline 
SKM1ad &1  &$ -4.0$ & 7.2 &19.6 \\
SKM4ad &4  &$ -3.4$ & 4.1 & 9.8 \\ 
SKM8ad &8  &$ -5.7$ & 6.1 & 6.8 \\ 
SKM16ad &15 &$-4.7$ &4.9 &5.1 \\ \hline \hline 
 \label{dynamonumbers}\end{tabular}}
Note. Runs marked with * are not dynamo active, hence the wavenumber of the growing dynamo
mode is extracted from other runs of similar aspect 
ratio. 
\end{table}

For incoherent $\alpha$-shear-driven dynamos,
the relevant dynamo number reads
\citep[see, e.g.,][]{BRRK08}
\begin{equation}
D_{\alpha S} = \frac{\alpha_{\rm rms} \left| S \right| }{\etaT^2 k_z^3},
\end{equation}
where usually only
the fluctuations of  $\alpha_{yy}$  are considered for $\alpha_{\rm rms}$.
They determined the critical $D_{\alpha S}$ to be 
$\approx 2.3$ for white-noise $\alpha$ 
fluctuations.
\cite{BRRK08} also reported that the diagonal and off-diagonal components of the $\alpha$ tensor were
nearly equal. In the SMHD cases studied here, this is no longer the case, as is shown
in \Fig{fig:histo}, where  
$\alpha_{xx}$
dominates.

\cite{BRRK08} also discussed the possibility of a contribution from an incoherent SC
effect by fluctuations of $\eta_{yx}$
with vanishing mean.
They studied a model where both incoherent
effects were acting together, the incoherent $\alpha$ effect  
mainly through  $\alpha_{yy}$
while the incoherent SC effect through $\eta_{yx}$
is described by the dynamo number 
\begin{equation}
D_{\eta_{\rm rms} S} = \frac{\eta_{yx,\rm rms} \left| S \right| }{\etaT^2 k_z^2}. \label{Eq:ISC}
\end{equation}
They found that for small $D_{\eta_{\rm rms} S}$ the critical
dynamo number, detected
for the incoherent $\alpha$ effect alone, was not much altered, 
while for higher values that critical number could be clearly reduced.
Hence, to decide which dynamo effect is at play in systems with large 
fluctuations, one should always  consider the dynamo numbers for
both incoherent effects simultaneously. 

Moreover, the presence of an additional coherent SC effect can alter the 
dynamo excitation condition, which we now account for by adding a 
term from a coherent $\eta_{yx}$
to the simplified zero-dimensional (0D) dynamo model of \cite{BRRK08}; see their Appendix~C. 
The equation solved is the linear mean-field induction equation
\EQA\label{MF}
\partial_t \overline{\bm{A}}=-S \meanA_y \hat{{\bm x}} + \meanEMF - \eta \meanJJ, \ENA
where the mean EMF now reads
\EQA\label{EMF_MF}
\meanemf_i=\alpha_{ij,\rm inc}(t) \meanB_j - \left( \eta_{yx,\rm inc}(t) + \eta_{yx}\right) \delta_{i2}  \meanJ_x.  
\ENA
The incoherent effects are modeled with $\delta$-correlated
noise in time having zero means and standard deviations equal to the respective rms values,
while the coherent contribution from $\eta_{yx}$ is constant.
By the ansatz $\meanAA \sim \exp(\ii k_z z)$, \eq{MF} turns into the 0D model,
with governing parameters  
$D_{\alpha S}$, $D_{\eta S}$, and $D_{\eta_{\rm rms} S}$,
defined above.

\begin{figure*}\begin{center}
\includegraphics[width=0.4\textwidth]{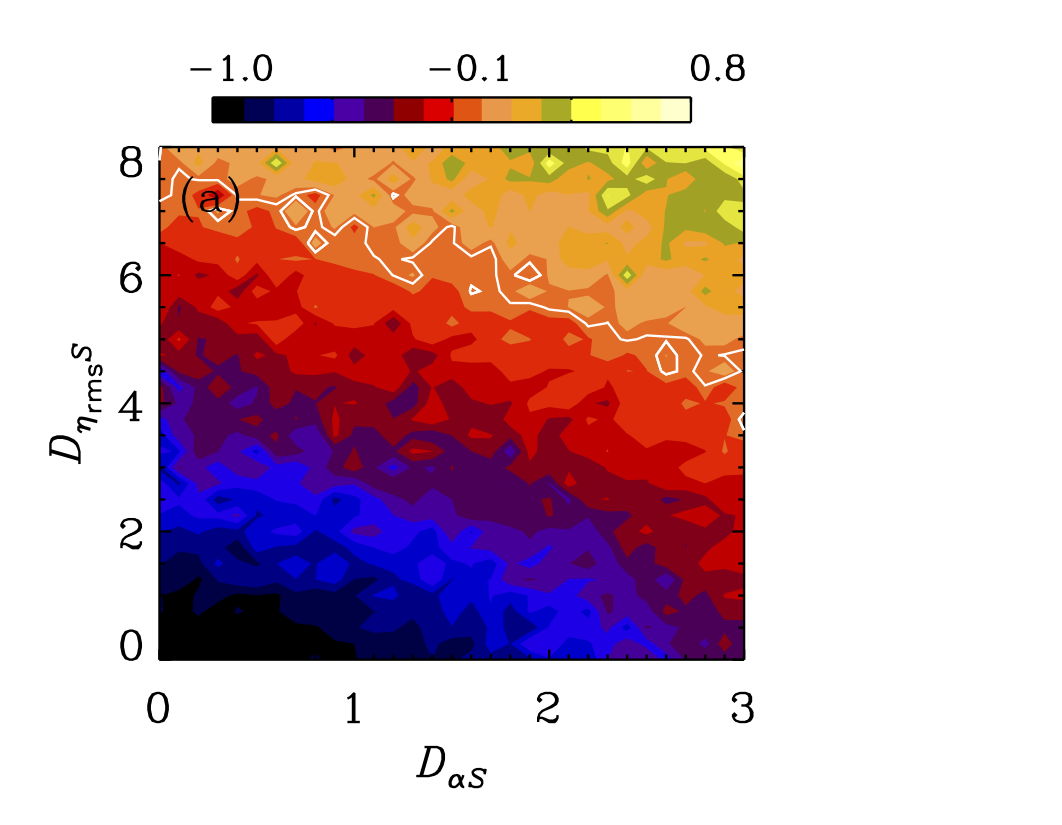}\hspace*{-1.5cm}
\includegraphics[width=0.4\textwidth]{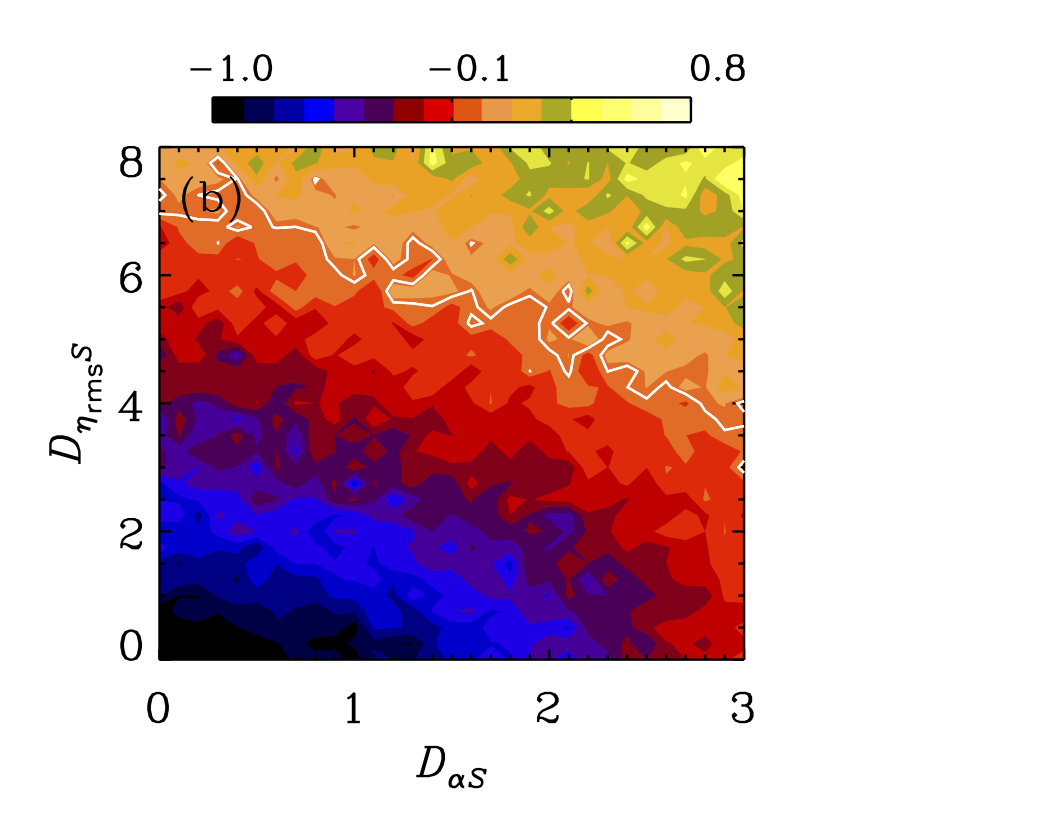}\hspace*{-1.5cm}
\includegraphics[width=0.4\textwidth]{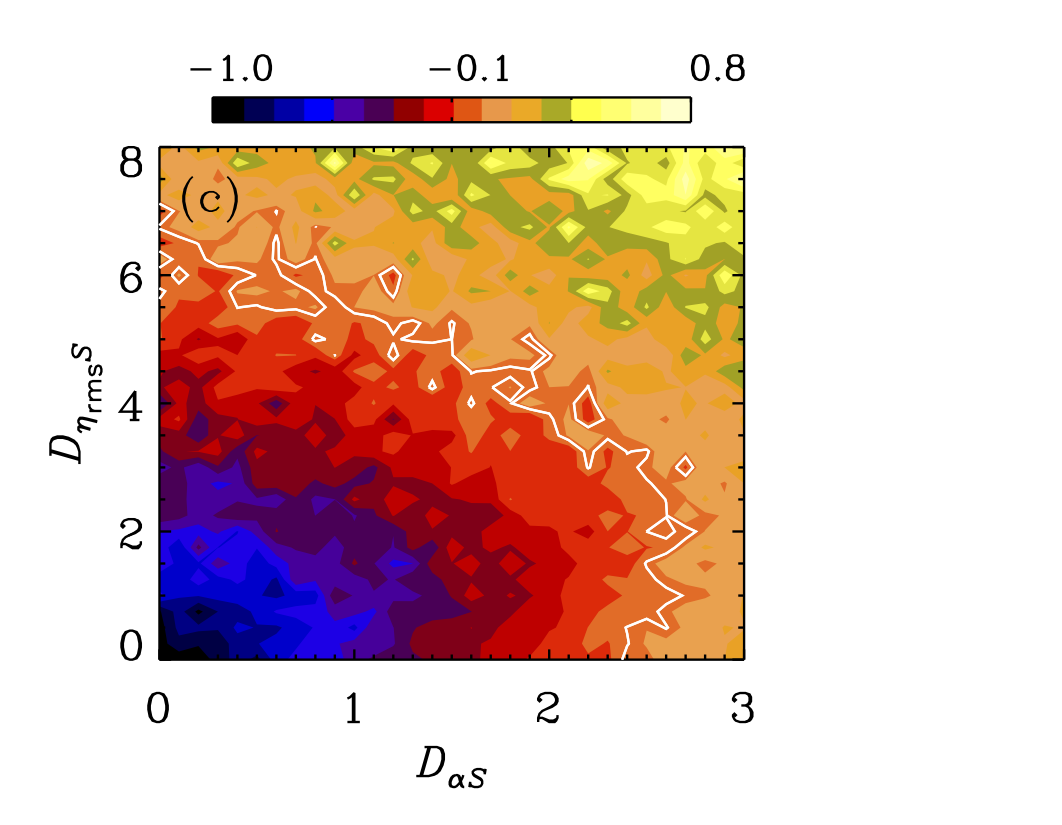}\\
\includegraphics[width=0.4\textwidth]{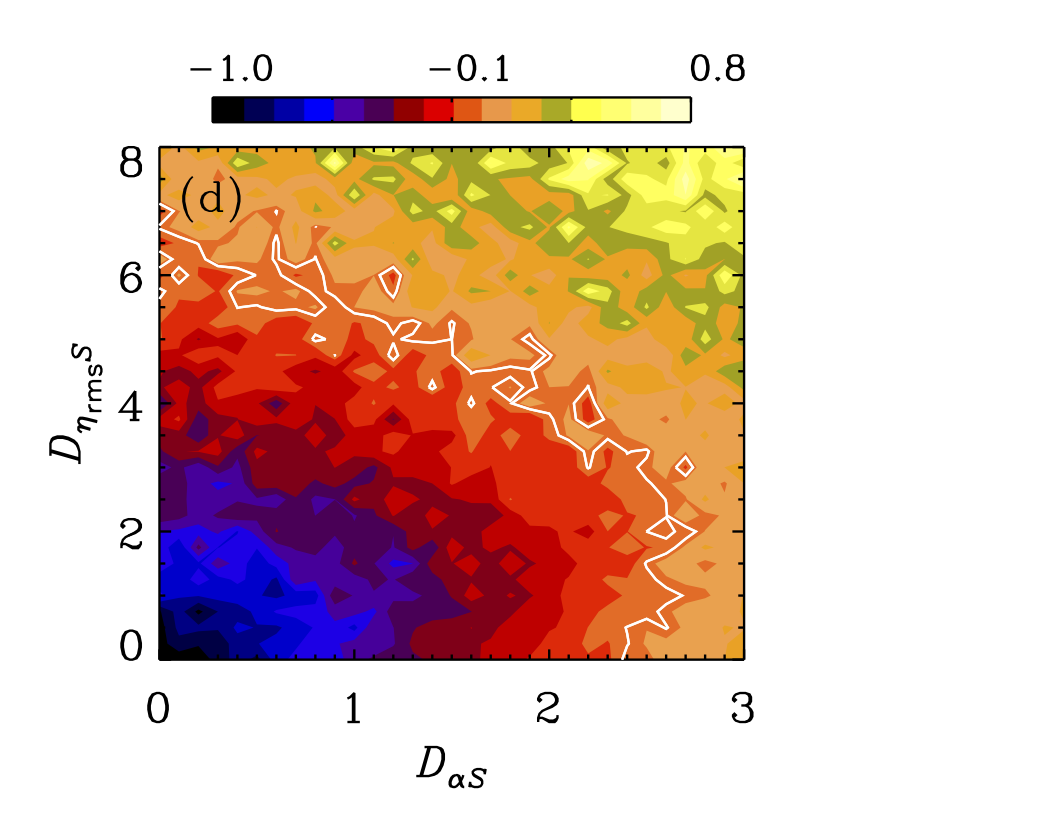}\hspace*{-1.5cm}
\includegraphics[width=0.4\textwidth]{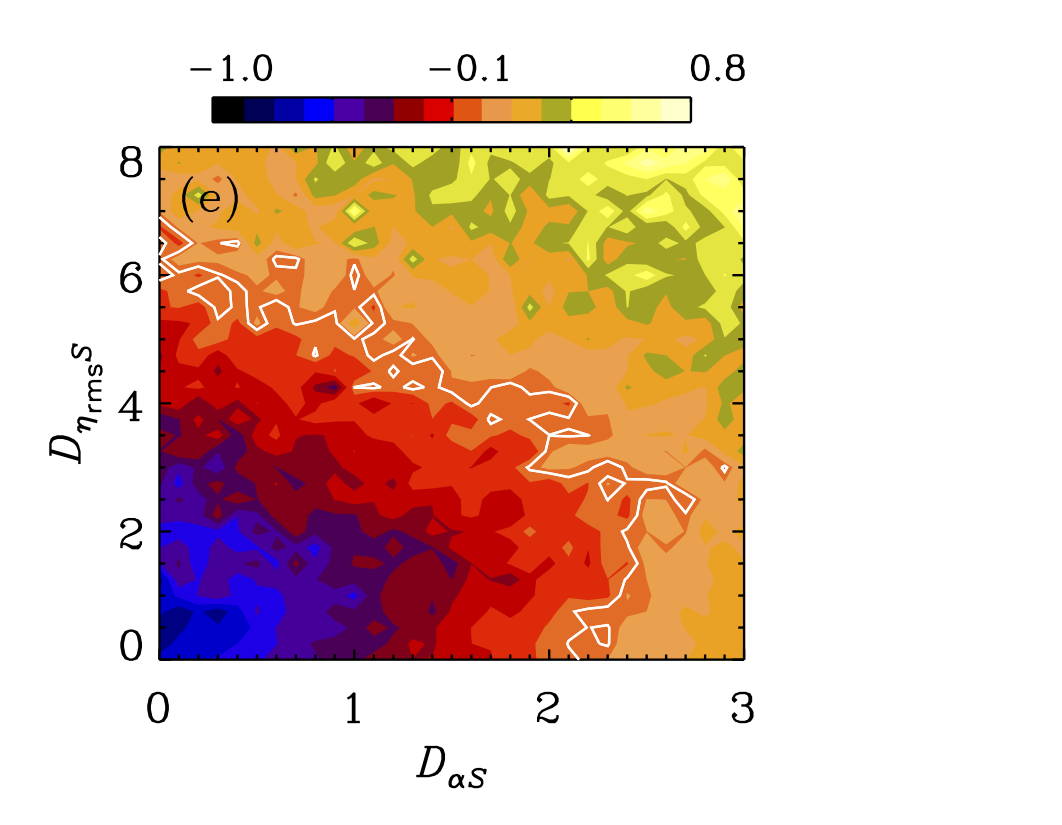}\hspace*{-1.5cm}
\includegraphics[width=0.4\textwidth]{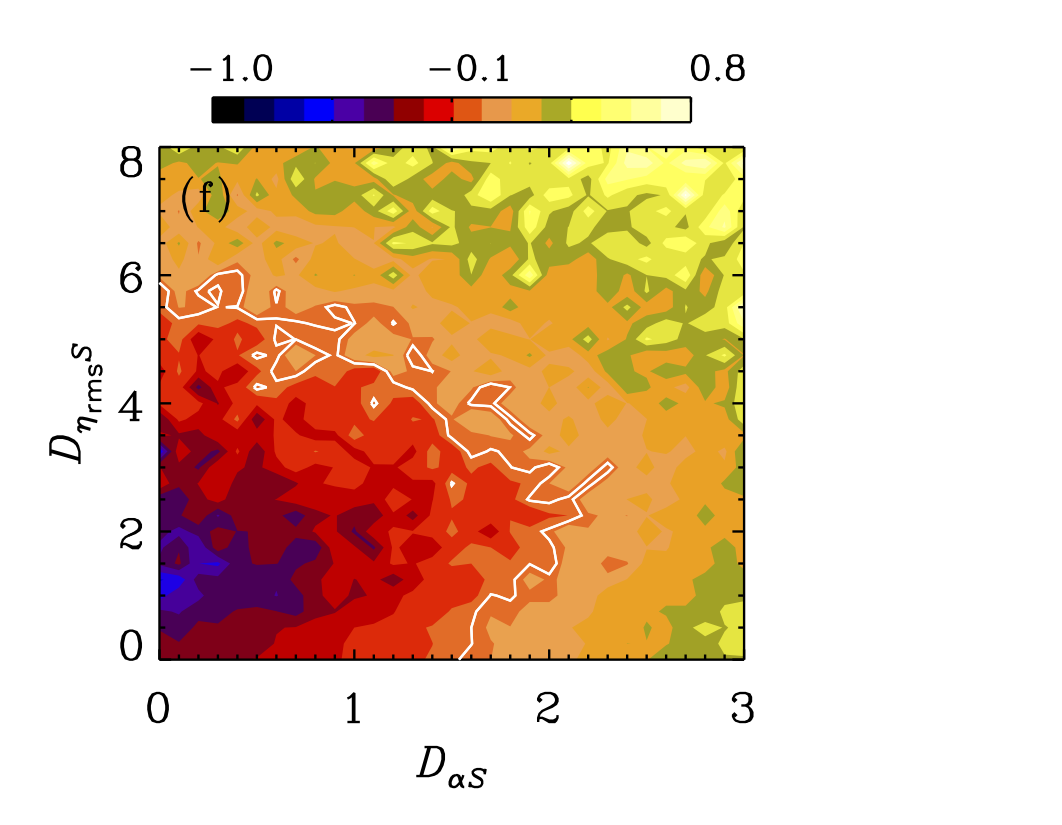}\\
\includegraphics[width=0.4\textwidth]{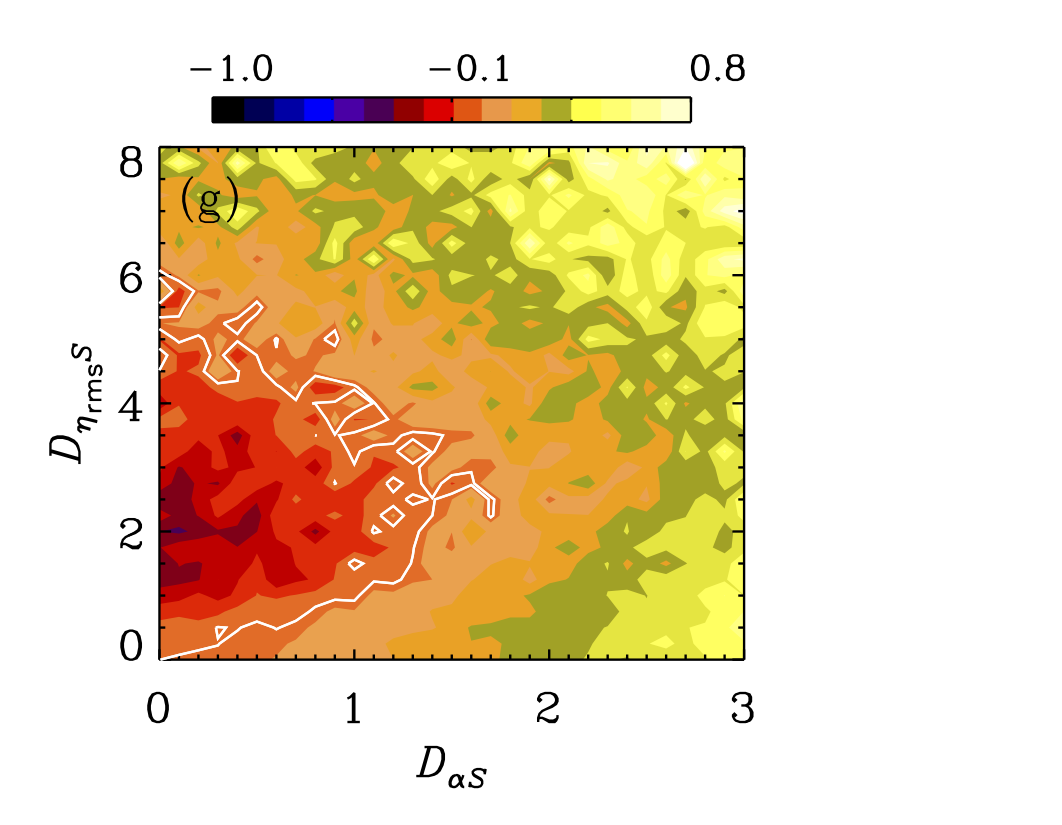}\hspace*{-1.5cm}
\includegraphics[width=0.4\textwidth]{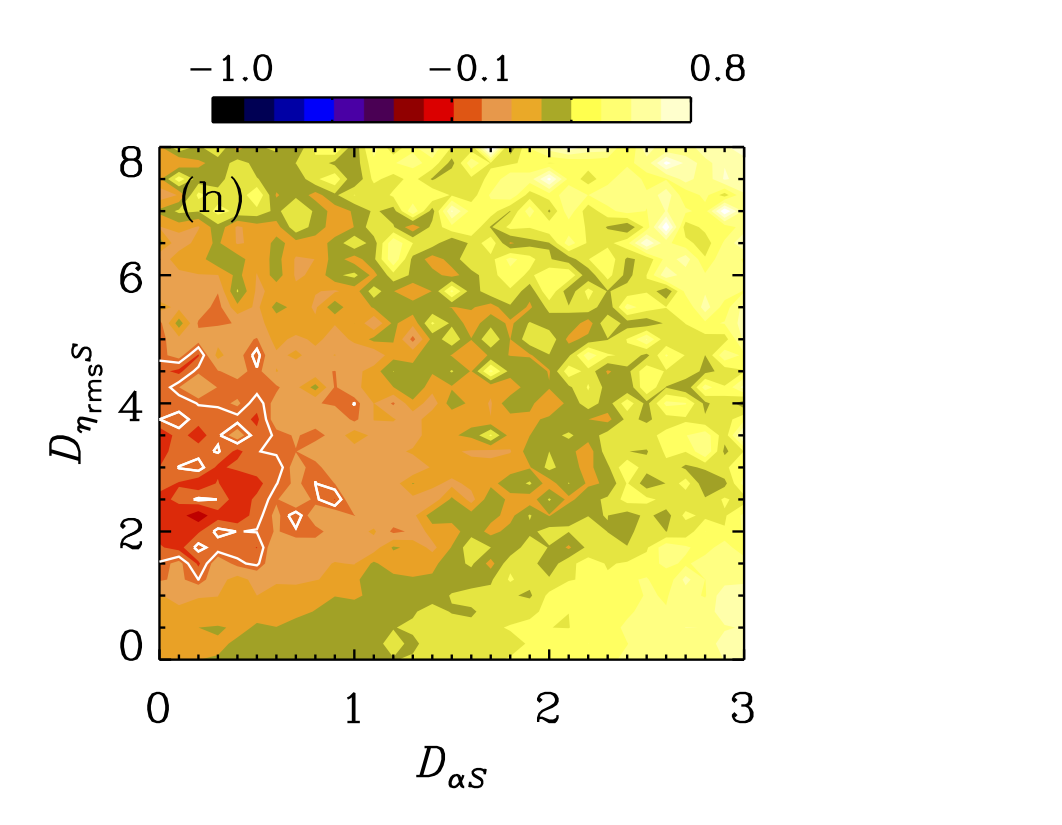}\hspace*{-1.5cm}
\includegraphics[width=0.4\textwidth]{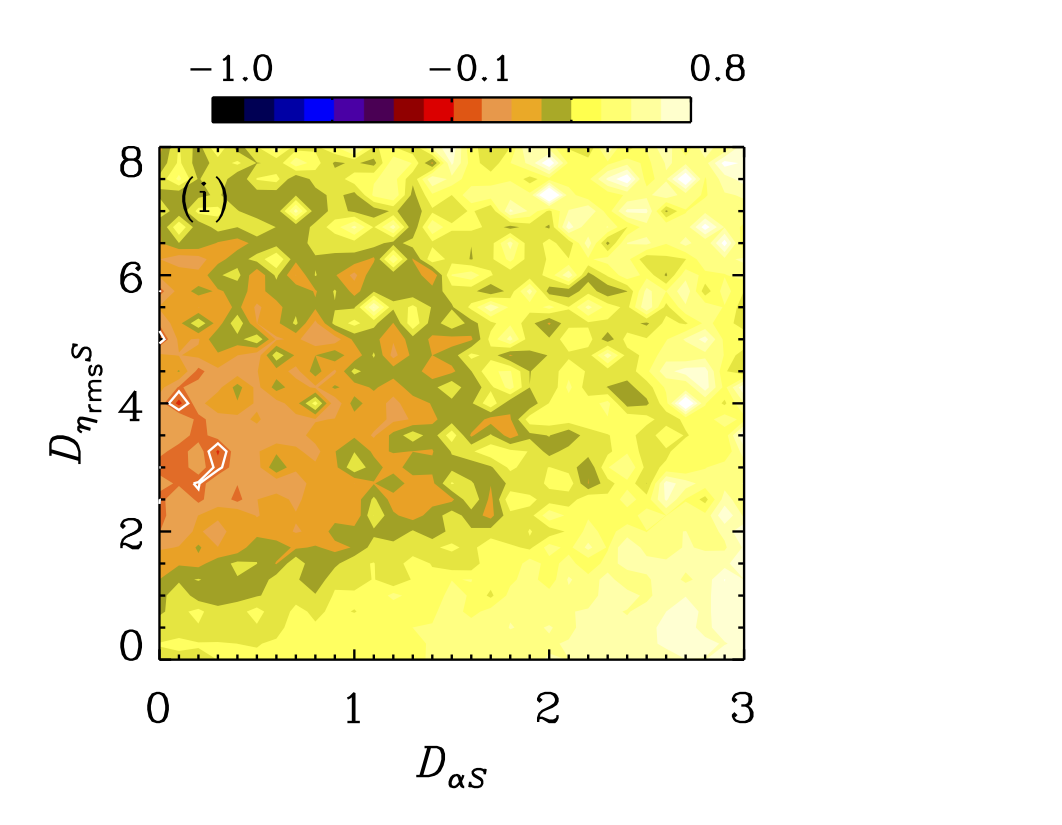}\\
\end{center}\caption{
Stability diagrams for different values of $D_{\eta S}$: from top left to bottom right, -1.5,-1.0,-0.5,-0.1,0.1,0.5,1.0,1.5, and 2.0.\\
White: zero-growth-rate contour.
Color scales: $\lambda/\etaT k_z^2$.
}\label{fig:stabilitydiagram}
\vspace{8mm}
\end{figure*}

We have verified that dynamo action in 
this
model without any incoherent
effects takes place when $D_{\eta S}$ is exceeding
unity, as expected from the stability criterion 
(\ref{Eq:SC}).
We compute new stability maps 
in the $D_{\eta_{\rm rms} S}\,$--$\,D_{\alpha S}$ plane for a series of dynamo numbers $D_{\eta S}$, 
in the range $[-1.5,2]$.
These values are similar in magnitude as those realized in our 
simulations, although not covering the extremal 
values obtained in the magneto-kinetic forcing cases.
These are shown in \Fig{fig:stabilitydiagram}, where 
panels (d) and (e) closely match the stability map
of the incoherent effects alone \citep[compare with
Figure~12 of][]{BRRK08}. As expected, adding a coherent SC effect with a positive 
$D_{\eta S}$ enhances the dynamo instability, especially by lowering the critical dynamo
number for the incoherent $\alpha$-shear dynamo. 
This is seen through the shift of the stability line 
(white contours in \Fig{fig:stabilitydiagram}) 
to the left (toward smaller values of $D_{\alpha S}$) from (f) to (i).
The incoherent SC dynamo threshold is also lowered,
but the effect is more subtle,
as seen through the much less dramatic shift of the stability boundary downward 
(toward smaller values of $D_{\eta_{\rm rms} S}$)
in Figure~\ref{fig:stabilitydiagram}, panels (f)--(i).
For $D_{\eta S} > 1$,
the coherent SC effect alone would result in the excitation of a dynamo, but the
presence of the incoherent effects causes small islands
in which dynamo action is suppressed;
see the dark red areas surrounded by the white contour in \Fig{fig:stabilitydiagram}, panels 
(g) and (h).   

In the dynamo numbers  (\ref{Eq:SC})--(\ref{Eq:ISC})
we also need the vertical wavenumber $k_z$ of the dynamo mode, which
we determined from Fourier analysis of the 
mean fields during the kinematic 
phase of the dynamo. For those runs that are not dynamo-active, we used $k_z$ from
a corresponding dynamo-active run with higher $\Rm$ 
(for kinetically forced runs) or a different forcing function (for  
magneto-kinetically forced runs),
but the same aspect ratio
(see Table~\ref{models}), and we denote those runs for which we obtained $k_z$
from elsewhere with an asterisk in Table~\ref{dynamonumbers}.
We also note that, if the dynamo enters saturation,
the kinematically preferred mode is not necessarily any longer present. Independent of the aspect 
ratio of the box, all the saturated models exhibit a magnetic field at the scale of the box
or, in other words, at the smallest permissible wavenumber.

In the FMHD cases, we obtain negative $D_{\eta S}$
and incoherent SC dynamo numbers of similar magnitude, 
with $D_{\alpha S}$ tending to be larger than $D_{\eta_{\rm rms} S}$,
especially in Run~FK1b.
In the case of Run~FK8a,
no dynamo action is seen, and none of the dynamo numbers predicts a dynamo either. In the
other case without dynamo, 
Run~FK1a, the $\eta$-related dynamo numbers
predict no dynamo action, while 
$D_{\alpha S}$ alone would
do so
($D_{\alpha S}=2.8>D_{\alpha S,\rm crit}=2.3$).
Its critical value, however, can be increased in 
this case, mainly by the presence of the rather strong 
coherent SC effect with a negative dynamo number.  
The two dynamo-active cases have 
$D_{\alpha S}$ clearly above the critical value.
Hence, the presence of moderate
suppressing factors cannot prevent the dynamo instability.
It clearly seems to
be the incoherent $\alpha$-shear one in the FMHD cases,
because $D_{\eta_{\rm rms} S}$ is far too small in this case.

In the kinetically forced SMHD cases,  however, $\eta_{yx}$ is negative, allowing
for the possibility of a coherent SC-effect
dynamo. All our runs of this type are dynamo-active, but only the high-$\Rm$ cases 
exhibit supercritical $D_{\eta S}$ ($>1$).
Except for the case of Run~SK4a,
the $D_{\alpha S}$ and $D_{\eta_{\rm rms} S}$ values
indicate supercriticality for 
the incoherent dynamo instabilities,
explaining again most of our findings. 
Run~SK4a has a low positive $D_{\eta S}$, but also the incoherent effects are well below
their critical dynamo numbers. The coherent SC effect could therefore assist the dynamo,
but this effect should be negligible according to the 0D model. Hence, this dynamo remains
unexplained with any dynamo scenario.
Dynamo excitation is easier In the SMHD models than in the 
FMHD ones, which might indicate that the coherent SC effect 
assists dynamo action, but 
the SMHD simplifications could also be the cause.

In the magneto-kinetically forced SMHD cases, the 
dynamo numbers indicate
stability against the MSC effect, but are all,
according to individual 0D model runs (not 
presented here),
supercritical for the incoherent dynamo effects,
the incoherent SC effect being even more pronounced now than in the kinetically
forced cases.
Although the cases with standard forcing do not show 
exponential growth, their decimated forcing counterparts do so.
Hence our interpretation here is
that a dynamo is present in all the cases with 
magneto-kinetic forcing. Even though
the coherent SC effect now exhibits larger negative dynamo numbers 
we find, by running individual 0-D models,
 that in all cases it should not be able to damp down the dynamo instability. 
Hence, again, the most likely mechanism for exciting the dynamo is the
incoherent $\alpha$-shear effect,
with supercritical $D_{\alpha S}$ in all cases. However, we cannot rule out the coexistence of 
an
incoherent SC effect, as some runs also indicate supercriticality against it.

\section{Conclusions}

We have studied different types of sheared MHD systems with the quasi-kinematic 
(QKTFM) and 
nonlinear (NLTFM) test-field methods.
In those cases studied with the NLTFM, we simplified the MHD equations
neglecting the pressure gradient in the 
momentum equation, which allows us to 
ignore the equation for the fluctuating density in the
test-field formulation, simplifying it somewhat.
In the case of the full MHD equations studied with the QKTFM, we
extend the previous results to even stronger shear, but still find no evidence for
negative values of the $\eta_{yx}$ component that could lead to LSD
action through the SC effect.

In kinetically forced magnetized burgulence (SMHD), we 
measure negative values of $\eta_{yx}$.
Indeed, dynamo action
with both radial and azimuthal magnetic field components growing exponentially
at the same rate is found.
The dynamo numbers for the coherent and the incoherent effects, based on the measured turbulent
transport coefficients, however, when employed in a simplified 0D dynamo model,
indicate that even in this case the dynamo is mainly driven by the
incoherent $\alpha$
effect and shear,
possibly assisted by the coherent SC effect.

In the case of systems with standard
magneto-kinetic forcing, we do not find 
exponential growth of the mean magnetic field.
When we repeat this experiment with a decimated forcing function, removing the 
smallest 
wavenumber
components,
exponential growth is recovered.
Hence, in our interpretation, there is still a dynamo instability in the magneto-kinetically forced cases,
but it becomes engulfed by the rapid growth of the mean field due the presence of
these low wavenumbers in the forcing, preventing us from seeing the exponential growth of the mean field.
The measured $\eta_{yx}$
are again positive and increasing as a function of the magnitude of
shear and the aspect ratio of the box,
and are therefore incapable of driving a dynamo through the MSC effect.
This finding is compatible with our analytical derivation predicting a positive
contribution to $\eta_{yx}$ in the case when the pressure term is neglected,
albeit restricted to ideal MHD; see \App{analyticetayx}.
The computed dynamo numbers, compared against the 0D model, again indicate
the most likely driver of the dynamo to be the incoherent $\alpha$
effect with shear.

We note that we have not investigated the magnetic 
Prandtl number ($\Pm$) dependence of the magneto-kinetically forced
cases although, according to the study of \cite{SB15a},
$\Pm$ has an influence on the magnitude of 
$\eta_{yx}$ (in their case always negative) 
such that its modulus decreases 
when $\Pm$ is increasing.
Their 
model
includes both rotation and shear, and in addition
they do not specify 
how $\Rm$ and $\Rey$  changed  when $\Pm$ was changed. Hence, the applicability of these results to our case is
uncertain, but studying the $\Pm$ dependence is an important future direction.
We also
acknowledge that the simplified MHD equations used here prevent
our conclusions from being generally applicable. Hence we cannot fully reject the
postulated possibility of a dynamo driven by the
MSC effect. 
The measurements
should be repeated with the full MHD equations, analyzed with a fully compressible
TFM, also solving for the density fluctuations. 

\begin{acknowledgements}
We acknowledge fruitful and inspiring discussions with Dr.\ Jonathan Squire and Prof.\
Amitava Bhattacharjee in the Max Planck Princeton Center for Plasma Physics framework.
M.J.K. and M.R.\ acknowledge the support of the Academy of Finland
ReSoLVE Centre of Excellence (grant number 307411).
This project has received funding from the European Research Council (ERC)
under the European Union's Horizon 2020 research and innovation
program (Project UniSDyn, grant agreement n:o 818665).
A.B.\ acknowledges supported through the Swedish Research Council,
grant 2019-04234, and the National Science Foundation under the grant
AAG-1615100.

\end{acknowledgements}

\bibliography{mara}{}
\bibliographystyle{aasjournal}

\appendix{
\section{Comparison of standard and decimated forcing} \label{sect:forcings}
\label{A}

To investigate the possible anisotropy due to the 
removal of all
 $|k_{x,y,z}|/k_1 \leq k_{\rm min}/k_1$ from the forcing
 (decimation),
we perform
two hydrodynamic simulations without shear.
Both runs were performed with $64^3$ grid points and $\kf/k_1=5$,
one without decimation and one with, using $k_{\rm min}/k_1=2$.
All other parameters were the same and $\urms$ was similar in the two cases,
with Mach number $\Ma=\urms/\cs=0.002$ and $\Rey=0.04$.
In \Fig{fig:pdfs_uu_decimated} we show probability density functions
(PDFs) of the three components of $\uu$ from
a snapshot of each run.
These PDFs are normalized such that $\int P(u_i)\, d u_i = 1$.
We find that the PDFs of $u_x$, $u_y$, and $u_z$ are 
in both cases on top of each other,
suggesting that the stochastic
flows are nearly isotropic, at least in a statistical sense.
Let us define the kurtosis, $\mbox{kurt}\,x$, of the distribution $P(x)$ as
\EQ
\mbox{kurt}\,x=\frac{1}{\sigma^4}\int_{-\infty}^{\infty}
(x-\overline{x})^4 P(x) d x - 3,
\EN
where $\overline{x}$ and $\sigma$ are its mean and variance,
respectively. We find that the kurtoses for all three velocity components
are nearly zero, suggesting Gaussian distributions.

Furthermore, we define a dimensionless
quantity $\zeta(\theta,\phi)=\sqrt{\bra{(\uu\cdot\nnn)^2}}/\urms$, useful to assess the degree of anisotropy,
with $\nnn=(\sin{\theta} \cos{\phi},\,\sin{\theta} \sin{\phi},\,\cos{\theta})$,
and the polar and azimuthal angles
$\theta$ and $\phi$, respectively,
as in a spherical coordinate system.
For the two runs discussed just above, we show in \Fig{fig:uuaniso}
distributions of $\zeta(\theta,\phi)$ that reveal
anisotropic features, both in the standard (undecimated)
and the decimated cases, at two different times.
However, at least in the undecimated case the flows
are expected to be
statistically isotropic when data from a large number of snapshots are
combined, as there is no preferred direction in the system.
We show the variation of $\zeta$ as a function of $\phi$ at two fixed values
of $\theta$ ($45\degr$ and $90\degr$) in \Fig{fig:uu_aniso_phi},
after performing an average over eight snapshots.
As expected, the degree of anisotropy decreased compared to 
a single snapshot; it is below $7\%$ as inferred from the values
of $\zeta$ in \Fig{fig:uu_aniso_phi}.
We also notice an azimuthal
 $m=2$ modulation which is more pronounced in the
decimated case, 
likely due to gaps in the thin 
$k$ shell around $\kf$.
The statistical isotropy of the
flow is expected to be  improved further at higher resolution and when   
data from a longer time series are combined. 

\begin{figure}\begin{center}
\includegraphics[width=\columnwidth]{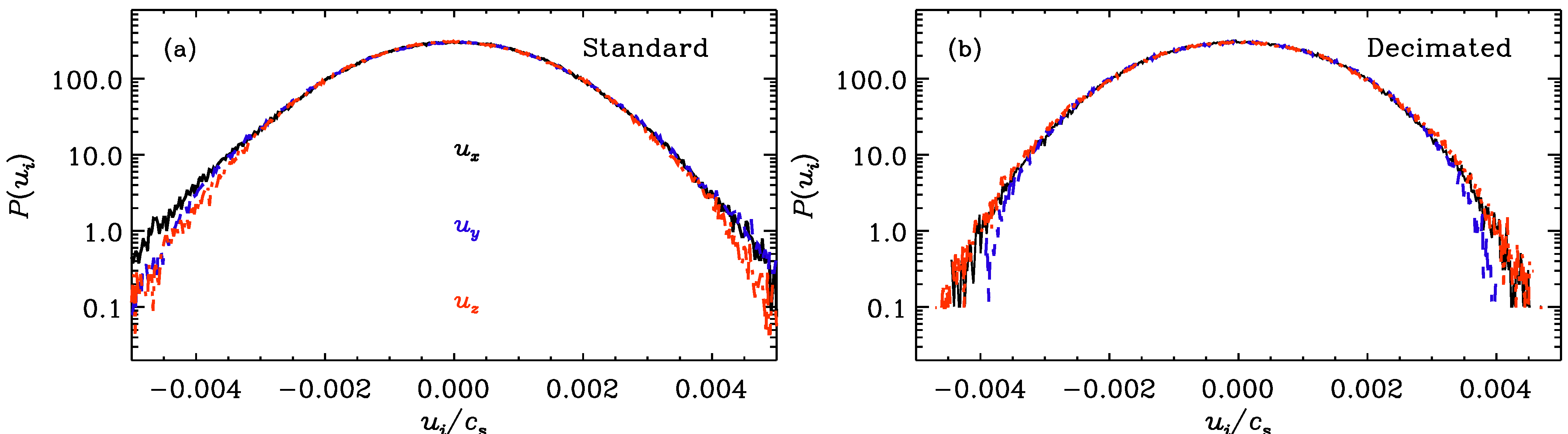}
\end{center}\caption{
PDFs of all three velocity components from $64^3$ shearless
hydrodynamic runs with $\kf/k_1=5$, discussed in \App{A}.
(a) standard, (b) decimated forcing with $k_{\rm min}/k_1=2$.
All pdfs are nearly Gaussians with kurtosis $\sim 0$.
}\label{fig:pdfs_uu_decimated}
\vspace{8mm}
\end{figure}

\begin{figure}\begin{center}
\includegraphics[width=\columnwidth]{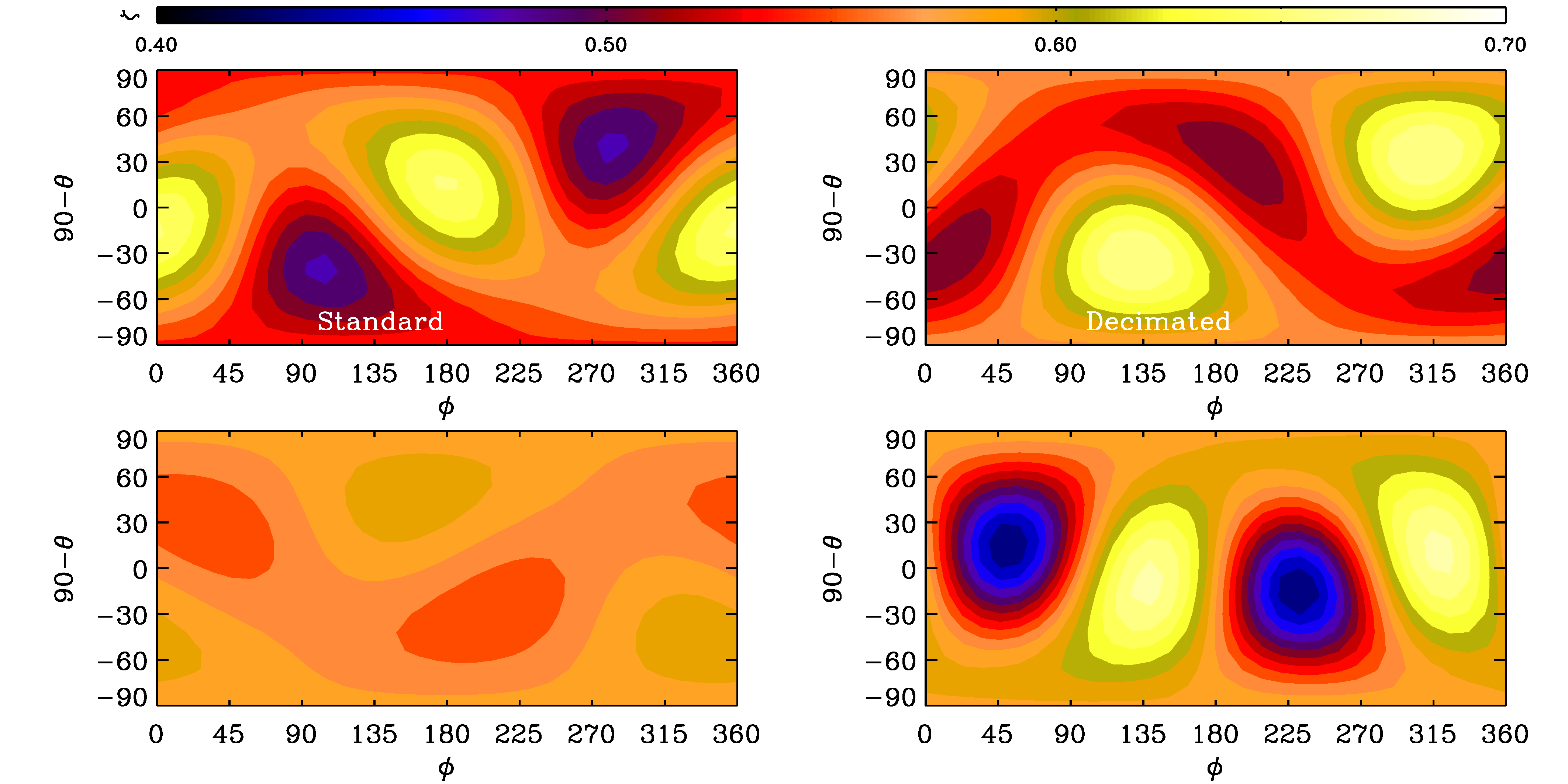}
\end{center}\caption{
Distribution of $\zeta=\sqrt{\bra{(\uu\cdot\nnn)^2}}/\urms$,
in the $\theta\phi$ plane at times $\nu \kf^2 t = 25$ (top) and
27.5 (bottom); from the two hydrodynamic runs discussed in \App{A}.
(a) Standard;
(b) decimated with $k_{\rm min}/k_1=2$.
}\label{fig:uuaniso}
\vspace{8mm}
\end{figure}

\begin{figure}\begin{center}
\includegraphics[width=0.45\columnwidth]{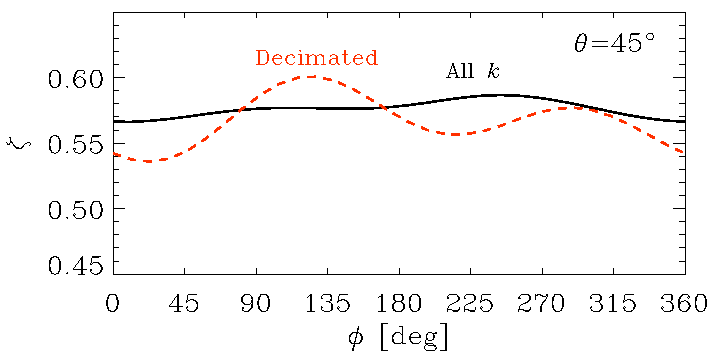}
\includegraphics[width=0.45\columnwidth]{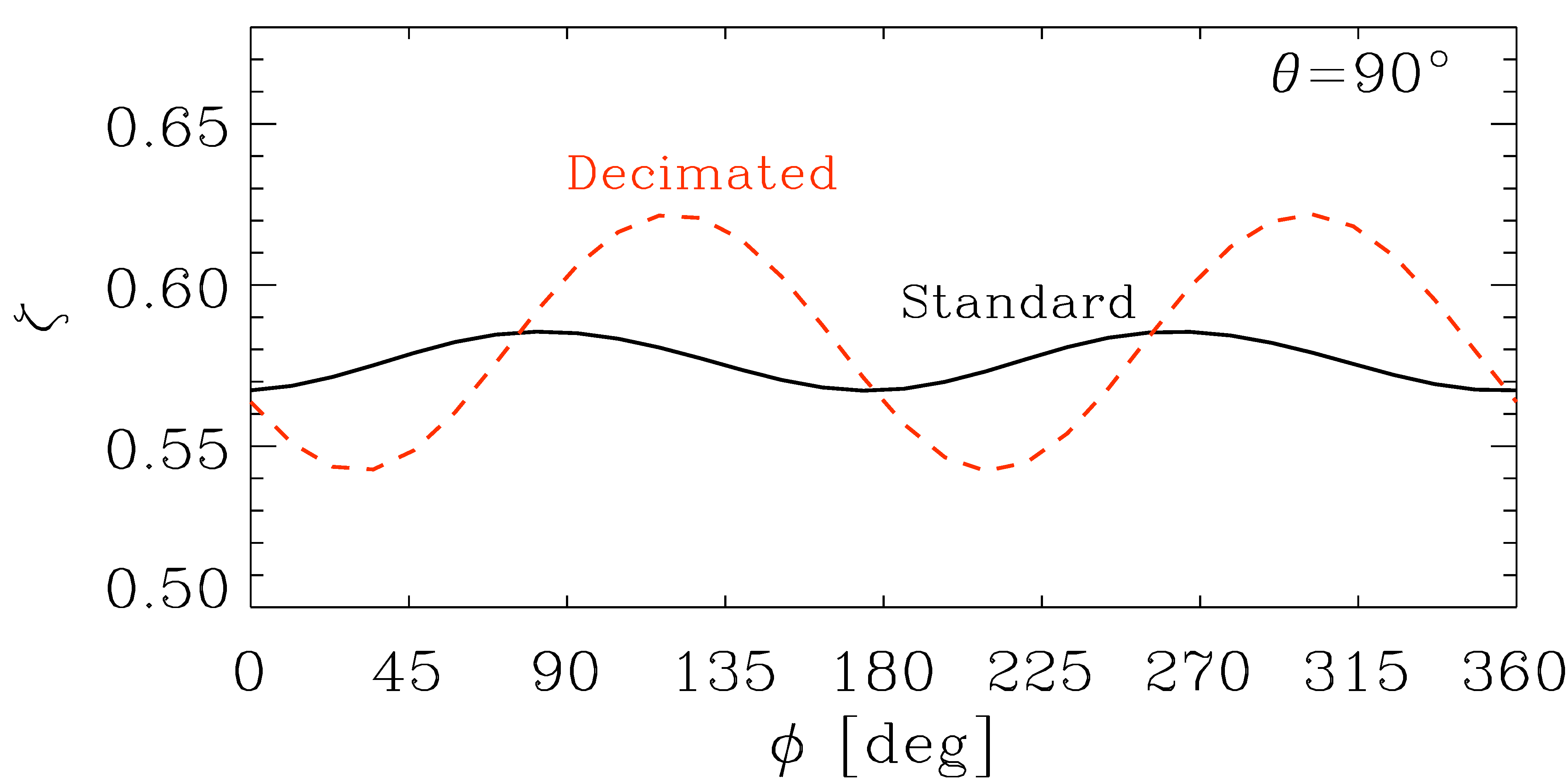}
\end{center}\caption{
Variation of $\zeta$
with the azimuthal angle $\phi$ at polar angles $\theta=45\degr$ (left)
and $90\degr$ (right), after averaging over eight snapshots from the
runs discussed in \App{A}.
Solid/black: standard;
dashed/red: decimated with $k_{\rm min}/k_1=2$.
}\label{fig:uu_aniso_phi}
\vspace{8mm}
\end{figure}

\section{Validation of the NLTFM} 

\subsection{Comparison of the different variants of the NLTFM}\label{sect:variants}

\begin{table}[t!]\caption{$\eta$ tensor components measured with the different variants of NLTFM
from Run~SKM1a007.}
\centerline{
\begin{tabular}{ccccc} \hline \hline
Method  & $\eta_{xx}/\eta_0$ & $\eta_{yy}/\eta_0$ & $\eta_{yx}/\eta_0$ & $\eta_{xy}/\eta_0$ \\ \hline
\hline
{\sf ju} &$2.110 \pm 0.023$  & $2.089 \pm 0.007$ & $0.112 \pm 0.026$ & $-0.208 \pm 0.028$ \\ 
{\sf jb} &$2.276 \pm 0.152$ &$2.106 \pm 0.020$ & $0.124 \pm 0.009$ & $-0.212 \pm 0.018$\\
{\sf bb} &$2.297 \pm 0.144$ & $2.116 \pm 0.018$ & $0.129 \pm 0.018 $& $-0.188 \pm 0.017$ \\
{\sf bu} & $2.155 \pm 0.047$ & $2.127 \pm 0.017$ & $0.113 \pm 0.014$ &$-0.212 \pm 0.022 $ \\ \hline \hline
\label{table:tfmvariants}\end{tabular}}
\end{table}

As is described in RB10, 
with respect to the terms $\uu\times\bb$ and $\jj\times\bb$
there are four possibilities to define the NLTFM, depending on 
how one combines
the fluctuating fields from the main run, $\uu$, $\bb$, $\jj$ with the test 
solutions  $\uu_{\meanB}$, $\bb_{\meanB}$, $\jj_{\meanB}$.
These variants
were denoted as 
{\sf ju} (using $\bm{j}$ and $\bm{u}$ in the pondero- and electromotive forces, respectively), 
{\sf jb} (using $\bm{j}$ and $\bm{b}$), 
{\sf bu} (using $\bm{b}$ and $\bm{u}$), and 
{\sf bb} (using $\bm{b}$ in both). 
Further variants due to the term $\uu\cdot\nab\uu$ are not considered here.
Previously it was concluded that the 
{\sf ju} method would be the most 
stable one 
(RB10).
Here we examine how 
the different variants 
behave in SMHD
with 
standard (random)
forcing.
The results for Run~SKM1a007 are listed in Table~\ref{table:tfmvariants} and
depicted in Figure~\ref{fig:tfm_variants},
showing the $\eta_{xx}$ component obtained with
all four variants.
We can see that {\sf jb} and {\sf bb} produce 
measurements that are nearly identical at any phase of the simulation.  
The measurements with {\sf bu}
deviate from these occasionally, 
but the largest deviations occur for {\sf ju}. 
While the three former variants
tend to produce turbulent transport coefficients that clearly grow 
within the resetting intervals, 
{\sf ju} produces plateaus, this 
difference being especially pronounced in Figure~\ref{fig:tfm_variants}, top panel. This is    
indicative of the test problems becoming unstable during the resetting interval, which can lead to 
 overestimation of and increased uncertainties in the measured 
transport coefficients. With the resetting time of 
$0.5 T_\nu$
in most of our simulations, however, the 
measured differences between the variants
were very small, but nevertheless we observed a tendency
 of the tensor components to be larger in magnitude when {\sf jb} and {\sf bb} were used;
see also Table~\ref{table:tfmvariants}.
 Hence, throughout the paper we use the {\sf ju} variant, which produces 
measurements with clearer plateaus in the turbulent transport coefficients.

\begin{figure*}\begin{center}
\includegraphics[width=0.8\textwidth]{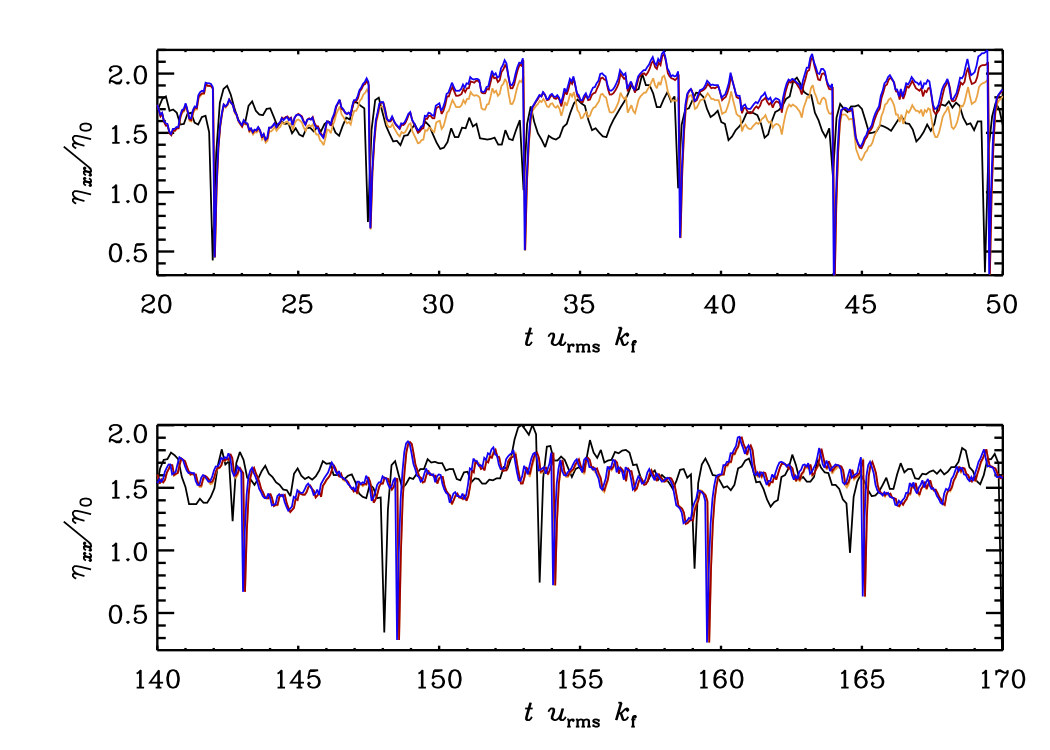}
\end{center}\caption{
Time evolution of 
$\eta_{xx}$ 
from Run~SKM1a007 with the four 
variants of the NLTFM. 
Black: {\sf ju}, blue: {\sf bb}, orange: {\sf bu}, red: {\sf jb}. Upper panel: early stages, 
lower panel: late stages of the simulation. 
Note that the {\sf jb} results are almost completely on top of the {\sf bu} ones.
}\label{fig:tfm_variants}
\vspace{8mm}
\end{figure*}

\subsection{Kinetically forced SMHD analyzed with QKTFM and NLTFM} \label{sect:comparison}

To further validate the NLTFM, we perform runs of kinetically forced SMHD,
and measure the turbulent transport coefficients with both QKTFM and NLTFM.
We compare them in two regimes: one where the magnetic field is very weak
and another
where the magnetic field is already dynamically significant. We choose
the setup SK4b, and show our results in Figure~\ref{fig:tfmskin} in terms of 
$\eta_{yx}$ as function of time.
 Although some differences due to the randomness of the forcing 
have to be expected, we observe a very good agreement between the two
methods.

\begin{figure*}\begin{center}
\includegraphics[width=\textwidth]{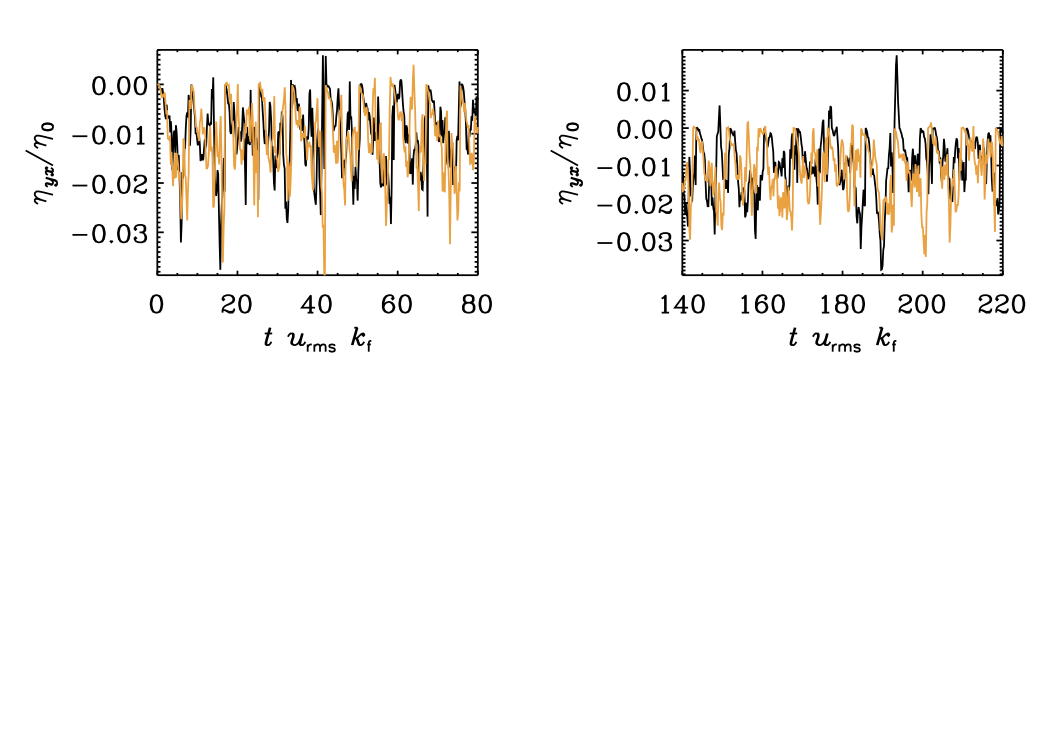}
\end{center}
\vspace{-7cm}
\caption{
Time evolution of 
$\eta_{yx}$
from Run~SK4b with QKTFM (orange) and NLTFM (black). Measurements are from a stage when the 
dynamo field
is still weak (left) and dynamically significant (right).
}\label{fig:tfmskin}
\end{figure*}

\section{An analytical estimate for $\eta_{\it{yx}}$}
\label{analyticetayx}

To obtain an analytical estimate for $\eta_{yx}$ in the
absence of the pressure term,
we assume ideal MHD ($\eta=\nu=0$) and constant density,
neglect terms quadratic in the fluctuations (SOCA),
 assume vanishing mean flow, except for $\UUS$, and
vanishing initial conditions of those parts of the fluctuations which are due to the influence of shear and $\meanBB$.
Then we have,
restricting the mean EMF to be of
first order in $\meanBB$,
 \begin{align}
     \eta_{yx} &= -\left\langle \bzz_x(y,t)\, I_1 -Sx \left( \bzz_x(y,t)  \,I_2 + \int_0^t \partial_y \bzz_x(\xi,\tau) d\tau \, I_1 \right) + \,S^2x^2  \int_0^t \partial_y \bzz_x(\xi,\tau) d\tau \, I_2 \right\rangle_{\!\!\!xy} \hspace{-0mm},\\ 
       I_1 &=\int_0^t \left(\bzz_y(\xi,\tau) + S \int_0^\tau \bzz_x(\xi',\tau') d\tau' \right) d\tau,\\
       I_2 &= \int_0^t \int_0^\tau \left( \partial_y \bzz_y(\xi',\tau') + S  \int_0^{\tau'}\partial_y \bzz_x (\xi{''},\tau{''}) d\tau{''} \right) d\tau' d\tau ,
\end{align}
where $\xi=y+Sx(\tau-t)$, $\xi'=y+Sx(\tau'-t)$ etc. 
and  the arguments $x$ and $z$ were dropped.
The magnetic field is in units of $\rho^{1/2}$ and $\left(\uu^{(00)},\bb^{(00)}\right)$ is the background turbulence (i.e., for $\meanBB=\boldsymbol{0}$) without influence of shear ($S=0$), hence
\begin{align}
  \partial_t \bb^{(00)} &= \nab\times\left(\uu^{(00)} \times \bb^{(00)} \right)' +\ffM \label{eq:backturb},\\
  \partial_t \uu^{(00)} &= - \left(\uu^{(00)}\cdot\nab \uu^{(00)} - \jj^{(00)}\times\bb^{(00)}\right)' + \ffK . \nonumber
\end{align}
Remarkably, there is no contribution from $\uu^{(00)}$ to $\eta_{yx}$
and only $\bzz_x\ne0$ is necessary for $\eta_{yx}\ne0$.
To zeroth and first order in $S$ we obtain
\begin{align}
\eta_{yx} = &- S \left\langle b_x^{(00)}(y,t)  \int_0^t \int_0^\tau b_x^{(00)}(y,\tau') d\tau' d\tau \right\rangle_{\!\!\!xy}  
                  - \left\langle b_x^{(00)} (y,t) \int_0^t b^{(00)}_y(y,\tau) d\tau\right \rangle_{\!\!\!xy} \label{eq:estim2} \\  
                   &-  S \left\langle x b_x^{(00)}(y,t)  \int_0^t \partial_y b^{(00)}_y(y,\tau)(\tau-t) \, d\tau \right\rangle_{\!\!\!xy}   
                   + S \left\langle x b_x^{(00)}(y,t)  \int_0^t \int_0^\tau \partial_y b^{(00)}_y(y,\tau') \, d\tau' d\tau \right\rangle_{\!\!\!xy}  \label{eq:estim4}\\
                  & + S \left\langle x  \left( \int_0^t \partial_y b^{(00)}_x(y,\tau) d\tau  \right)  \left(  \int_0^t b^{(00)}_y(y,\tau) d\tau \right)  \right\rangle_{\!\!\!xy},   \label{eq:estim5} 
\end{align}
where the second contribution in \eqref{eq:estim2} vanishes in isotropic background turbulence because of $\left\langle b_i^{(00)}(y,t) b_j^{(00)} (y,\tau) \right\rangle_{\!\!xy} \propto \delta_{ij}$.
Integration by parts in \eqref{eq:estim4} yields
\begin{align}
       \int_0^t \partial_y b^{(00)}_y(y,\tau)(\tau-t) \, d\tau = - \int_0^t \int_0^\tau \partial_y b^{(00)}_y(y,\tau') \, d\tau' d\tau.
\end{align}
Further, we have
\begin{align}
     b_x^{(00)}(y,t)  \int_0^t \int_0^\tau \partial_y b^{(00)}_y(y,\tau') \, d\tau' d\tau  = &\,\partial_y \left( b_x^{(00)}(y,t)  \int_0^t \int_0^\tau b^{(00)}_y(y,\tau') \, d\tau' d\tau \right) \\
      &- \partial_y b_x^{(00)}(y,t)  \int_0^t \int_0^\tau b^{(00)}_y(y,\tau') \, d\tau' d\tau, \label{eq:estim3}
\end{align}
in which the first term on the right vanishes under averaging over $y$. 
Hence, for the terms \eqref{eq:estim4} and \eqref{eq:estim5} we obtain
\begin{align}
    S \left\langle x \left[ \left( \int_0^t \partial_y b^{(00)}_x(y,\tau) d\tau  \right)  \left(  \int_0^t b^{(00)}_y(y,\tau) d\tau \right) - 2 \, \partial_y b_x^{(00)}(y,t)  \int_0^t \int_0^\tau b^{(00)}_y(y,\tau') \, d\tau' d\tau \right]\right\rangle_{\!\!\!xy}.
     \label{eq:estim6} 
\end{align}
Because of isotropy and mirror symmetry of the background turbulence, the correlator $\left\langle b_i^{(00)}(y,t) \partial_k b^{(00)}_j(y,\tau) \right\rangle_{\!\!xy}$ vanishes $\forall i,j$. Hence, it is only the factor $x$ in 
\eqref{eq:estim6} that possibly prevents this term from vanishing, in contrast to the first term in \eqref{eq:estim2}, which is based on a  correlator, usually assumed positive definite. 

On the other hand, truly Galilean-invariant turbulence should not exhibit an explicit $x$ dependence. Given that  the forcing in our simulations  indeed obeys Galilean invariance, deviations from it in $\uu$ and $\bb$ can only emerge
due to the ``memory'' of the turbulence, which is made ``everlasting" by the absence of dissipative damping.
Thus, for the purpose of interpreting our numerical results, we may disregard \eqref{eq:estim6} and assume that only the first term in \eqref{eq:estim2} determines the sign of $\eta_{yx}$. 

Comparing with \cite{SB15a} by setting $\nu=\eta=0$ in their Equations~(32)--(35)%
\footnote{Note that they employ a sign-inverted definition of $S$.}
we find the following agreements:
\begin{itemize}\itemsep-1pt

\item[1.] no contribution from $\uu^{(00)}$ (or $W_u$ in their terms) to $\eta_{yx}$, only from $\bb^{(00)}$ (or $W_b$),
\item[2.]
        $\eta_{yx}$ has the opposite sign of $S$ and is thus unfavorable for the MSC-effect dynamo \\
        (for this we have to assume a positive correlation $\left\langle b_x^{(00)}(y,t;x,z)\, b_x^{(00)}(y,\tau';x,z)  \right\rangle_{\!\!xy}$).
\end{itemize}

In summary, our analytical result is in
qualitative
 accordance with the numerical ones for the magneto-kinetic forcing
cases. 
For the purely kinetic ones, however, the analytics predicts 
vanishing
$\eta_{yx}$, 
while the numerical experiments do produce 
it,
even with a favorable sign for dynamo action, albeit too weak to be 
its main driver,
and also weaker than in comparable magneto-kinetic forcing setups.
As 
vanishing $\eta_{yx}$
is in agreement with the ideal limit of \cite{SB15a}, 
we conclude that 
a nonzero contribution 
to $\eta_{yx}$ from kinetic fluctuations and shear (their $(\eta_{yx})_u^S$),
requires the presence of
dissipative terms, most likely $\eta\ne0$, as their result (32) 
suggests.
It also reveals that there must be an ``optimal" magnitude of $\eta$ that maximizes 
$\left|(\eta_{yx})_u^S\right|$
because it
vanishes again in the limit $\eta\rightarrow\infty$. To be too far from the optimal $\eta$ in numerical setups might explain the absence of a dynamo, enabled by 
$(\eta_{yx})_u^S$.

\end{document}